\documentclass[12pt]{article}
\usepackage[english]{babel}
\usepackage{amsmath,amsthm}
\usepackage{amsfonts}
\usepackage{graphicx}
\usepackage{enumerate}
\usepackage[usenames,dvipsnames]{color}
\usepackage{subfigure}
\usepackage[right,pagewise,displaymath, mathlines]{lineno}
\usepackage{epstopdf}
\usepackage{color}
\usepackage{graphicx}
\usepackage{times}
\usepackage{latexsym}
\usepackage{epsfig}
\usepackage{natbib}
\usepackage{enumitem}

\numberwithin{equation}{section}
\numberwithin{figure}{section}
\numberwithin{table}{section}

\numberwithin{equation}{section}

\def\be{\mathbf{e}}
\def\bE{\mathbf{E}}
\def\bI{\mathbf{I}}
\def\bL{\mathbf{L}}
\def\bS{\mathbf{S}}
\def\bu{\mathbf{u}}
\def\bU{\mathbf{U}}
\def\bx{\mathbf{x}}
\def\bX{\mathbf{X}}
\def\bz{\mathbf{z}}
\def\bZ{\mathbf{Z}}

\def\bgamma{\mbox{\boldmath $\gamma$}}
\def\bLambda{\mbox{\boldmath $\Lambda$}}
\def\bmu{\mbox{\boldmath $\mu$}}
\def\bpi{\mbox{\boldmath $\pi$}}
\def\bPsi{\mbox{\boldmath $\Psi$}}
\def\bSigma{\mbox{\boldmath $\Sigma$}}
\def\btheta{\mbox{\boldmath $\theta$}}
\def\bTheta{\mbox{\boldmath $\Theta$}}
\def\bXi{\mbox{\boldmath $\Xi$}}

\def\cL{\mathcal{L}}
\def\cN{\mathcal{N}}

\def\EE{\mathbb{E}}

\begin{document}


\begin{center}
{\Large 
Robust estimation for mixtures of Gaussian factor analyzers, based on trimming and constraints}
\vspace*{8mm}

{\small
L.A. Garc\'{i}a-Escudero$^{\textrm{a}}$,
A. Gordaliza$^{\textrm{a}}$,
F. Greselin$^{\textrm{b}}$,
 S. Ingrassia$^{\textrm{c}}$,
 A. Mayo-Iscar$^{\textrm{a}}$}
\bigskip

{\scriptsize
$^{\textrm{a}}$\textit{Department of Statistics and  Operations Research and IMUVA, University of Valladolid, Valladolid, Spain}
\smallskip

$^{\textrm{b}}$\textit{Department of Statistics and Quantitative Methods, Milano-Bicocca University, Milan,  Italy}
\smallskip

$^{\textrm{c}}$\textit{Department of Economics and Business, University of Catania, Catania, Italy}
}
\end{center}
\medskip

\begin{quote}
\noindent
\textbf{Abstract.} Mixtures of Gaussian factors are powerful tools for modeling an
unobserved heterogeneous population, offering  - at the same time -
dimension reduction and  model-based clustering. Unfortunately, the
high prevalence of spurious solutions and the disturbing effects of
outlying observations, along maximum likelihood estimation, open
serious issues. In this paper we consider restrictions
for the component covariances, to avoid spurious solutions, and
trimming, to provide robustness against violations of normality
assumptions of the underlying latent factors. A detailed AECM
algorithm for this new approach is presented. Simulation results and an application to the AIS dataset show the
aim and effectiveness of the proposed methodology.

\medskip

\end{quote}

{\footnotesize 
\tableofcontents 
}

\section{Introduction and motivation}
Factor analysis is an effective method  of summarizing the variability between a number of correlated features, through a much smaller number of unobservable, hence named \textit{latent}, factors. It originated from the consideration that, in many phenomena, several observed variables could be explained by a few unobserved ones. Under this approach, each single observed variable (among the $p$ ones)  is assumed to be a linear combination of $d$ underlying common factors with an accompanying error term to account for that part of variability that is unique to it (not in common with other variables). Ideally, $d$ should be substantially smaller than $p$, to achieve  parsimony.

Clearly, the effectiveness of this method is limited by its global linearity, as it happens for principal components analysis. Hence,  \citet{Ghah:Hilt:1997}, \citet{Tipp:Bish:1997} and  \citet{McLaPe:MFA:2000b} solidly widened  the  applicability of these approaches  by combining local models of Gaussian factors in the form of finite mixtures. 
   The idea is to employ latent variables to perform dimensional reduction in each component, thus providing a statistical method which concurrently performs clustering and, within each cluster, local dimensionality reduction.

In the literature, error and factors are routinely assumed to have a Gaussian distribution because of their mathematical and computational tractability: however statistical methods which ignore departure from normality may cause biased or misleading inference. Moreover, it is well known that maximum likelihood estimation for mixtures often leads to ill-posed problems because of the unboundedness of the objective function to be maximized, which favors the appearance of non-interesting local maximizers and degenerate or \textit{spurious} solutions.

The lack of robustness in mixture fitting arises whenever the sample
contains a certain proportion of data that does not follow the
underlying population model. Spurious solutions can even appear when
ML estimation is applied to artificial data drawn from a given
finite mixture model, i.e. without adding any kind of contamination.
Hence, robustified estimation is needed. Many contributions in this
sense can be found in the literature: from Mclust model with a noise
component in \cite{Fral:Raft:Howm:1998}, mixtures of
$t$-distributions in \cite{McLa:Peel:Robu:1998}, the trimmed
likelihood mixture fitting method in \cite{NeyF07}, 
 the trimmed ML estimation of contaminated
mixtures in \cite{GalR09}, and the robust improper ML estimator
introduced in \cite{coretto2011maximum}, 
among many others. Some important
applications in fields like computer vision, pattern recognition,
analysis of microarray gene expression data, or tomography (see, for
example, \cite{stewart1999robust}, \cite{Campbell19971539},
\cite{Bickel01052003} and \cite{maitra2001clustering}, respectively)
suggest that more attention should be paid to robustness, because
noise in the data sets may be frequent in all these fields of
application.

Different types of constraints have been traditionally applied in Gaussian mixtures of factor analyzers, 
for instance, some authors propose to take a common (diagonal)
error matrix (as for the Mixtures of Common Factor Analyzers,
denoted by MCFA, in \citealp{Baek:McLa:MixtFA:2010}) or to impose an
isotropic error matrix \citealp{Bish:Tipp:1998}. This strategy has
proven to be effective in many cases, at the expenses of stronger
distributional restrictions on the data. In
\citealp{McNi:Murp:Pars:2008}, when analyzing parsimonious mixtures
of Gaussians factor analyzers models, they realized that equal
determinant restrictions give more stable results. For avoiding
singularities and spurious solutions, under milder conditions,
\cite{Gres:Ingr:Maxi:2013} recently proposed to maximize the
likelihood by constraining the eigenvalues of the covariance
matrices, following previous work  of \citet{Ingr:2004} and going
back to \citet{Hath:Acon:1985}. Furthermore, \citet{McLa:Bean:2005},
\citet{Baek:McLa:MixtCT:2011}, \citet{steane2012model} and
\citet{lin2014capturing} have considered the use of mixtures of
$t$-analyzers in an attempt to make the model less sensitive to
outliers, but they, too, are not robust against very extreme
outliers \citep{Henn:2004}; while \citet{fokoue2003mixtures}
proposed a Bayesian approach.

The  purpose of the present work  is to  introduce an
estimating procedure for mixture of Gaussian factors analyzers that
can resist the effect of outliers  and avoid spurious
local maximizers. The proposed constraints can be also used to take
into account prior information about the scatter parameters. 

Trimming has been shown to be a simple, powerful, flexible and
computationally feasible way to provide robustness in many different
statistical frameworks. The basic idea behind trimming here is the
removal 
of a little proportion $\alpha$ of observations whose values would
be the more unlikely to occur if the fitted model was true.  
In this way,  trimming avoids 
that a small fraction of outlying
observations could exert a harmful effect on the estimation of the
parameters of the fitted model.

Incorporating constraints in the mixture fitting estimation method
moves the mathematical problem in a well-posed setting and hence
minimizes the risk of incurring spurious solutions. Moreover, a
correct statement of the  problem  allows to study the properties of
the EM algorithms as in \cite{Ingr:Rocc:2007} and to obtain the
desired statistical properties for the estimators, such as the
existence and consistency results,  as in \cite{GarG08} and
\cite{GalR09}.

We have organized the rest of the paper as follows. In Section
\ref{sec:GaussianMFA} we introduce notation and summarize main ideas
about Gaussian Mixtures of Factor Analyzers (in the foremost,
denoted by MFA). Then, in Section \ref{sec:TrimmedMFA} we introduce
the trimmed likelihood for MFA and we provide fairly extensive notes
concerning  the  EM algorithm, with incorporated trimming and
constrained estimation.  
In Section \ref{sec:Numerical results} we discuss  the  performance
of our procedure, on the ground of some numerical results obtained
from simulated and real data. In particular, we compare the bias and
MSE of robustly estimated model parameters for different  cases of
data contamination, by Monte Carlo experiments. The application to
the Australian Institute of Sports dataset shows how classification
and factor analysis can be developed using the new model. Section
\ref{sec:Concluding} contains concluding notes and provides ideas
for further research.

\section{Gaussian Mixtures of Factor Analyzers}\label{sec:GaussianMFA}
%
The density of the $p$-dimensional random variable $\bX$ of interest is modeled as a
mixture of $G$ multivariate normal densities in some unknown proportions $\pi_1,\ldots \pi_G$, whenever each data point is taken to be a realization of the following density function,
\begin{equation}
f(\bx;\btheta)=\sum_{g=1}^G \pi_g \phi_p(\bx;\bmu_g,\bSigma_g)\label{mixt-Gaussian}
\end{equation}
where $\phi_p(\bx;\bmu,\bSigma)$ denotes the $p$-variate normal density function with mean vector $\bmu$ and covariance matrix $\bSigma$. Here the vector $\btheta=\btheta_{GM}(p,G)$ of unknown parameters consists of the $(G-1)$ mixing proportions $\pi_g$, the $G p$ elements of the component means $\bmu_g$, and the $\frac{1} {2} G p (p+1)  $ distinct elements of the component-covariance matrices $\bSigma_g$.
MFA postulates a finite mixture of linear sub-models for the distribution of the full observation vector $\bX$, given the (unobservable) factors $\bU$. That is, MFA provides local dimensionality reduction  by assuming that the distribution of the observation  $\bX_i$  can be given as
\begin{equation}
\bX_i=\bmu_g+\bLambda_g\bU_{ig}+\be_{ig} \quad\textrm{with probability }\quad \pi_g \: (g=1,\ldots,G) \quad \textrm{for} \,\, i=1,\ldots,n, \label{factor_an}
\end{equation}
where $\bLambda_g$ is a $p \times d$ matrix of \textit{factor loadings}, the  \textit{factors} $\bU_{1g},\ldots, \bU_{ng}$ are $\cN(\mathbf{0},\bI_d)$ distributed independently of the  \textit{errors} $\be_{ig}$. The latter are  independently $\cN(\mathbf{0},\bPsi_g)$  distributed, and $\bPsi_g$ is a $p \times p$ diagonal matrix $(g=1,\ldots,G)$.  The diagonality of $\bPsi_g$ is one of the key assumptions of factor analysis: the observed variables are independent given the factors. Note that the factor variables $\bU_{ig}$ model correlations between the elements of $\bX_i$, while the errors $\be_{ig}$  account for independent noise for $\bX_i$.   We suppose that $d<p$, which means that  $d$ unobservable factors are jointly explaining the $p$ observable features of the statistical units.
Under these assumptions, the mixture of factor analyzers model is given by (\ref{mixt-Gaussian}), where the $g$-th component-covariance matrix $\bSigma_g$ has the form
\begin{equation}
\bSigma_g=\bLambda_g \bLambda'_g+\bPsi_g \quad (g=1,\ldots,G). \label{Sigmag}
\end{equation}
The parameter vector $\btheta=\btheta_{MFA}(p,d,G)$ now consists of
the elements of the component means $\bmu_g$, the $\bLambda_g$, and
the $\bPsi_g$, along with the mixing proportions $\pi_g$
$(g=1,\ldots,G-1)$, on putting $\pi_G=1-\sum_{i=1}^{G-1}\pi_g$.


\section{Trimmed  Mixtures of  Factor Analyzers}\label{sec:TrimmedMFA}

In this section we present the  \textit{trimmed (Gaussian) mixtures of  factor
 analyzers model}  (trimmed MFA) and we propose a feasible
algorithm for its  implementation.

\subsection{Problem statement}\label{se3_1}

We will fit a mixture of Gaussian factor components to a given
dataset $\bx=\{\bx_1,\bx_2,\ldots,\bx_n\}$ in $\mathbb{R}^p$ by
maximizing a \textit{trimmed mixture log-likelihood} (see
\citealt{NeyF07}, \citealt{GalR09} and \citealt{GarG:ADAC:2014})
defined as: 
\begin{equation}\label{d1}
  \cL_{trim}=\sum_{i=1}^n z(\bx_i) \log\left[\sum_{g=1}^G \phi_p
(\bx_i ;\mu_g,\bLambda_g \bLambda_g' + \bPsi_g)\pi_g \right]
\end{equation}
where
$z(\cdot)$ is a 0-1 trimming indicator function that tell us whether
observation $\bx_i$ is trimmed off: $z(\bx_i)$=0, or not:
$z(\bx_i)$=1 and $\bSigma_g=\bLambda_g \bLambda_g^{'}+\bPsi_g$ as in
(\ref{Sigmag}). A fixed fraction $\alpha$ of observations can be
unassigned by setting $\sum_{i=1}^n z(\bx_i)=[n(1-\alpha)]$. Hence
the parameter
$\alpha$ denotes the trimming level. 
As usual, $\bx_1,\ldots,\bx_n$ are the realized values of $n$ independent  and identically distributed  random vectors $\bX_1,\ldots,\bX_n$ with common density given in (\ref{mixt-Gaussian}), with  component-covariance matrices $\bSigma_g$  as in  (\ref{Sigmag}) for $g =1,\ldots,G$. The component label vectors $\bz_1,\ldots,\bz_n$ are taken to be the realized values of the random vectors $\bZ_1,\ldots,\bZ_n$, where, for independent feature data, it is appropriate to assume that they are  (unconditionally) multinomially distributed. i.e. $\bZ_1,\ldots,\bZ_n \sim^{i.i.d.} Mult_G(1;\pi_1,...,\pi_G)$.\\

Moreover, to avoid the unboundedness of $\cL_{trim}$, we introduce a
\textit{constrained maximization} of (\ref{d1}). In more detail,
with reference to the 
diagonal elements $\{\psi_{g,k}\}_{k=1,...,p}$ of the
noise matrices $\bPsi_g$ for $g=1,\ldots,G$ we require that
\begin{equation}
    \psi_{g_1,k}\leq c_{noise} \,\,\, \psi_{g_2,h} \quad \quad \mbox{ for every }1 \leq k  \neq h \leq p \mbox{ and }1 \leq g_1 \neq g_2 \leq G \label{constraintOnPsi}
\end{equation}

The constant $c_{noise}$ is finite and
such that $c_{noise} \geq 1 $, to  avoid the $|\Sigma_g| \rightarrow 0$ case.  This constraint can be seen
as an adaptation to MFA of those introduced in \citet{Ingr:Rocc:2007},
\citet{GarG08} and it is similar to the mild  restrictions  implemented for MFA in \cite{Gres:Ingr:Maxi:2013}. They all go back to the seminal paper of
\citet{Hath:Acon:1985}.
We will look for the ML estimators of $\bPsi_g$ under the given constraints, and this position set  the maximization problem as a well-defined one, and at the same time discard singularities and  reduce spurious solutions.

If $\{\lambda_k(A)\}_{k=1,...,p}$ denote the set of
eigenvalues of the
$p\times p$ matrix $A$, a second set of constraints apply on 
the product of the loading matrices $\bLambda_g
\bLambda'_g$ by requiring that
\begin{equation}
    \lambda_{k}(\bLambda_{g_1}\bLambda'_{g_1})\leq c_{load} \,\,\, \lambda_{h}(\bLambda_{g_2}\bLambda'_{g_2})
    \mbox{ for every }1 \leq k  \neq h \leq d \mbox{ and }1 \leq g_1 \neq g_2 \leq G. \label{constraintOnLambda}
\end{equation}
with $c_{load}$ such that $1 \leq c_{load} < +\infty$.
These $\lambda_{k}(\bLambda_{g}\bLambda'_{g})$ values
control the different scatters in the reduced subspaces.
In fact, these type of constraints are not needed to avoid
singularities in the target function but they could be useful to
achieve more sensible solutions.


In the foremost, we will denote by  $\Theta_{c}$ the constrained
parameter space for $\btheta=\{\pi_g,\allowbreak \mu_g,\allowbreak
\Psi_g,\allowbreak \Lambda_g; \allowbreak g=1,\ldots,G\}$ under  the
requirements (\ref{constraintOnPsi}) and (\ref{constraintOnLambda}).

\subsection{Algorithm}\label{se3_2}

The  maximization of $\cL_{trim}$ in (\ref{d1}) for $\theta \in \Theta_{c}$ is  not an easy
task, obviously. We will give a feasible algorithm  obtained
by combining the Alternating Expectation-Conditional Maximization algorithm (AECM) for MFA with that (with trimming and
constraints) introduced in \cite{GarG:ADAC:2014} \citep[see,
also,][]{FriG13}.

The AECM is an extension of the EM, suggested by the factor structure of the model, that uses different specifications of missing data at each stage. The idea is to partition the vector of parameters $\btheta=(\btheta'_1,\btheta'_2)'$ in such a way that  $\cL_{trim}$ is easy to be maximized for $\btheta_1$ given $\btheta_2$ and viceversa, replacing the M-step   by a number of computationally simpler conditional maximization (CM) steps. In more detail, in the first cycle we set $\btheta_1=\{\pi_g, \bmu_g; g=1,\ldots,G\}$ and the missing data are the unobserved  group labels $\bZ=(\bz'_1,\ldots,\bz'_n)$, while  in the second cycle we set  $\btheta_2=\{\bLambda_g, \bPsi_g;g=1,\ldots,G\}$ and the missing data are the   group labels $\bZ$ and the unobserved latent factors $\bU=(\bU_{11},\ldots,\bU_{nG})$. Hence, the application of the AECM algorithm consists of two cycles, and there is one E-step and one CM-step alternatively considering  $\btheta_1$ and $\btheta_2$ in each cycle. Before describing the algorithm, we remark that the unobserved group labels  $\bZ$ are considered missing data in both cycles. Therefore, during the $l$-th iteration, we shall denote by $z_{ig}^{(l+1/2)}$ and $z_{ig}^{(l+1)}$ the conditional expectations at the first and second cycle, respectively. \\

\vspace{0.7cm}
 The algorithm has to be run multiple times on the same dataset, with different starting values, to prevent the attainment of a local, rather than global, maximum log-likelihood. In each run it executes the following steps:
\begin{enumerate}
  \item[1] \textit{Initialization:}\\
  Each iteration begins by  selecting
   initial values for $\theta^{(0)}$ where $\theta^{(0)}=(\pi_g^{(0)},\allowbreak \mu_g^{(0)},\allowbreak
 \bLambda_g^{(0)},\allowbreak \bPsi_g^{(0)}; g=1,\ldots,G)$.
  Inspired from results obtained in a series of extensive test experiments about  initialization strategies \citep[see][]{maitra2009initializing},   and aiming to allow
  the algorithm to visit the entire parameter space, we randomly select $p+1$ units  (without replacement) for  group $g$  from the observed data $\bx=\{\bx_i\}_{i=1,\ldots,n}$. In this way we
  obtain a subsample $\bX^g$ that we arrange in a $(p+1)\times p$ matrix, and its
  sample mean will be the initial $\mu_g^{(0)}$.
 Additionally, based on these $p+1$ observations, we developed a new \textit{ad hoc} approach for providing an initialization procedure
 for $\bPsi_g^{(0)}$ and $\bLambda_g^{(0)}$, to deal with the possible existence of gross outlying observations among the subsamples,
 which could inflate disproportionally some of their eigenvalues.
 The rationale under our procedure is, as usual, to fill in randomly the missing information in the complete model through random subsamples and,
 then, to estimate the other parameters. The missing information here are 
 the factors $\bu_1,\ldots,\bu_n$,
 which, under the assumptions for the model, are  independently $\cN(\mathbf{0}, \bI_d)$  distributed.
 We consider  model (\ref{factor_an}) in group $g$ as  a regression of $\bX_i$ with intercept $\bmu_g$,
 regression coefficients given by $\bLambda_g$, where the explanatory variables are the latent factors $\bU_{ig}$, and with regression errors $\be_{ig}$. Hence
 we draw $p+1$ random observations from the $d$-variate standard Gaussian to fill a  $(p+1) \times d$ matrix $\bU^g$. 
 Then we set  $\bLambda_g^{(0)}=(\bU_g' \bU_g)^{-1} \bU_g'\bX^g_c$ where $\bX^g_c$ is obtained by centering the columns of the $\bX^g$
 matrix. 
To provide a restricted random generation of $\bPsi_g$,  we
 compute the $(p+1)\times p$ matrix
$\mathbf{\varepsilon}_g=\bx_c^g-\bLambda_g^{(0)} \bU_g$, and we set
the diagonal elements of $\bPsi_g^{(0)}$ equal to the variances of
the $p$ columns of the $\mathbf{\varepsilon}_g$ matrix. We repeat this for $g=1,\ldots,G$ and if the obtained matrices
$\bLambda_g^{(0)}$ and $\bPsi_g^{(0)}$ do not satisfy the required
constraints  (\ref{constraintOnPsi}) and (\ref{constraintOnLambda}),
then the constrained maximizations described in step 2.4 must be
applied.
  Finally,
  weights $\pi_1^{(0)},...,\pi_G^{(0)}$ in the interval $(0,1)$ and summing up to 1 are
  randomly chosen.

\vspace{0.7cm}
  \item[2] \textit{Trimmed AECM steps:}\\
  The following steps 2.1--2.4. are alternatively executed until convergence
  (i.e. $||\bz^{(l+1)}-\bz^{(l)}||< \epsilon$)  for a small constant $\epsilon
>0$  or until reaching a maximum number of iterations $MaxIter$. The implementation of trimming is related to the  ``concentration" steps applied in high-breakdown robust methods \citep{RouD98}. Trimming is performed along the E-steps, while constraints are enforced during the second cycle CM step.

\begin{enumerate}
  \vspace{0.7cm}
    \item[2.1] \textit{First cycle. E-step:} \\  Here $\btheta_1=\{\pi_g, \bmu_g; g=1,\ldots,G\}$ and the missing data are the unobserved  group labels $\bz=(\bz'_1,\ldots,\bz'_n)$. The E-step on the first cycle on the $(l+1)$-th iteration requires the calculation of
$$Q_1\left(\btheta_1; \btheta^{(l)}\right) = \EE_{\btheta^{(l)}} \big[\cL_{trim} (\btheta_1) | \bx \big],$$
which is the  expected trimmed complete-data log-likelihood given the data $\bx$ and using the current estimate $\btheta^{(l)}$ for $\btheta$. In practice it requires calculating
$\EE_{\btheta^{(l)}} [Z_{ig}| \bx ]$ and usual computations show that this step is achieved by replacing each $z_{ig}$ by its current conditional expectation given the observed data $\bx_i$, that is we replace $z_{ig}$ by $\tau_{ig}^{(l+1/2)}$, where the latter is evaluated as follows.
 Let us define
   $$
    D_{g}\left(\bx;\theta^{(l)}\right)=\phi_p \left( \bx; \bmu_g^{(l)} , \bLambda_g^{(l)} {\big[\bLambda_g^{(l)}\big]}' + \bPsi_g^{(l)} \right) \pi_g^{(l)}
    $$
    and
   $$
    D_i=D\left(\bx_i;\theta^{(l)}\right)= \sum_{g=1}^G D_{g}\left(\bx_i;\theta^{(l)}\right), \mbox{ for }i=1,...,n.
    $$
    After sorting these $n$ values, the notation
    $
    D_{(1)} \leq.... \leq
    D_{(n)}
    $
    is adopted. Let us consider the subset of indices $I \subset\{1,2,...,n\}$ defined
    as
      \begin{equation}
      I=\big\{i : D_{(i)}\geq D_{([n \alpha])}   \big\}.\label{eq:D(i)}
     \end{equation}

    To update the parameters, only the observations with indices in $I$ will be taken into account. In other words,
     we are tentatively discarding the proportion $\alpha$ of observations with
    the smallest $D_{(i)}$ values.

   Then, set
$$
  \tau_{ig}^{(l+1/2)} = \left\{
  \renewcommand{\arraystretch}{2}
        \begin{array}{ll}
        \displaystyle{
        \frac{D_{g}(\bx_i;\theta^{(l)})}{D(\bx_i;\theta^{(l)})}} &  \mbox{ for }i\in I \\
        0  &  \mbox{ for }i\notin I. \\
               \end{array}
                    \right.
$$

Note that, for the observations with indices in $I$,  $\tau_{ig}^{(l+1/2)}$ are the ``posterior probabilities" often
    considered in standard EM algorithms applied when fitting MFAs. But, unlike the standard EM algorithms,  the
    $\tau_{ig}^{(l+1/2)}$ (and consequently the $z_{ig}$) for the discarded observations are
    set to 0.

  \vspace{0.7cm}
    \item[2.2] \textit{First cycle. CM-step:} This first CM step requires the maximization of $Q_1(\btheta_1; \btheta^{(l)})$ over $\btheta$, with $\btheta_2$ held fixed at $\btheta_2^{(l)}$.  We get $\btheta_1^{(l+1)}$  by updating  $\pi_g$ and $\mu_g$ as follows
    $$
    \pi_g^{(l+1)}= \frac{\sum_{i=1}^n \tau_{ig}^{(l+1/2)}}{[n(1-\alpha)]}
    $$
    and
    $$
    \mu_g^{(l+1)}= \frac{\sum_{i=1}^n \tau_{ig}^{(l+1/2)}\bx_i }{ n_g^{(l+1/2)}}
    $$

where $n_g^{(l+1/2)}= \sum_{i=1}^n \tau_{ig}^{(l+1/2)}$, for
$g=1,\ldots,G$. According to notation in \cite{McLa:Peel:fini:2000},
we set $\btheta^{(l+1/2)}=\left(\btheta_1^{(l+1)},\btheta_2^{(l)}\right)'$.

      \vspace{0.7cm}
      \item[2.3] \textit{Second cycle. E- step:} \\
      Here we consider $\btheta_2=\{  (\bLambda_g$, $\bPsi_g), \,  g=1, \ldots, G \}$, where the missing data are the unobserved group labels $\bZ$ and the latent factors $\bU$.
Therefore, the trimmed complete-data  log-likelihood in this second
cycle may be written as
\begin{equation*}
\cL_{trim:2}(\btheta_2)  =\sum_{i=1}^n  z(\bx_i) \log \sum_{g=1}^G \left[ \phi_p \left( \bx_i ;\bmu_g^{(l+1)}-\bLambda_g\bu_{ig}, \bPsi_g \right) \phi_d \left( \bu_{ig};0,\bI_d \right) \pi_g^{(l+1)} \right]. \nonumber \\
\label{eq:complete-data likelihood2}
\end{equation*}

%
The E-step on the second cycle on the $l$-th iteration requires the
calculation of  the conditional expectation of $\cL_{trim:2}$, given
the observed data $\bx$ and using  the current estimate
$\btheta^{(l+1/2)}$ for $\btheta$, i.e. $$Q_2\left(\btheta_2;
\btheta^{(l+1/2)}\right) = \EE_{\btheta^{(l+1/2)}}\big[\cL_{trim:2}(\btheta_2)
| \bx \big].$$

In addition to updating  the posterior probabilities  $\EE_{\btheta^{(l+1/2)}} [Z_{ig}| \bx ]$ by performing a concentration step and replacing each $z_{ig}$ by the new values $z_{ig}^{(l+1)}=\tau_{ig}^{(l+1)}$ (and consequently $n_g^{(l+1)}= \sum_{i=1}^n   \tau_{ig}^{(l+1)}$, for $g=1,\ldots,G$, as previously done in step 2.1),
this leads to evaluate the following conditional expectations:  $\EE_{\btheta^{(l+1/2)}}[ Z_{ig} \bU_{ig} | \bx] $
and $ \EE_{\btheta^{(l+1/2)}} [ Z_{ig}\bU_{ig} \bU_{ig}' | \bx ]$.
Recalling that the conditional distribution of $\bU_{ig}$ given $\bx_i$ is
\begin{align*}
\bU_{ig} | \bx_i \sim \cN \left( \bgamma_g(\bx_i-\bmu_g),\bI_q-\bgamma_g\bLambda_g \right)
\end{align*}
for $i=1,\ldots,n$ and $g=1,\ldots,G$ with
\begin{align*}
\bgamma_g=\bLambda_g' (\bLambda_g \bLambda_g'+\bPsi_g)^{-1},
\end{align*}
we obtain
\begin{align*}
\EE_{\btheta^{(l+1/2)}} [ Z_{ig} \bU_{ig} | \bx_i ] &=z_{ig}^{(l+1)}\bgamma_g^{(l)} \left( \bx_i- \bmu_g^{(l+1)} \right) \\
\EE_{\btheta^{(l+1/2)}} [Z_{ig} \bU_{ig} \bU_{ig}' | \bx_i ]&=z_{ig}^{(l+1)} \bigg[ \bgamma_g^{(l)} \left( \bx_i-\bmu_g^{(l+1)} \right) \left( \bx_i-\bmu_g^{(l+1)} \right)' {\bgamma_g^{(l)}}'  + \bI_q-\bgamma_g^{(l)} \bLambda^{(l)}_g \bigg] \\
&= z_{ig}^{(l+1)} \bXi^{(l)}_{ig},
\end{align*}
where we set
\begin{align*}
\bgamma^{(l)}_g & ={\bLambda_g^{(l)}}' \left( \bLambda_g^{(l)}{\bLambda_g^{(l)}}'+\bPsi^{(l)}_g \right)^{-1} \\
\bXi^{(l)}_{ig} & =\bI_q-\bgamma_g^{(l)}
\bLambda_g^{(l)} +\bgamma_g^{(l+1)}
\left( \bx_i-\bmu_g^{(l+1)} \right) \left( \bx_i-\bmu_g^{(l+1)} \right)' {\bgamma_g^{(l)}}'.
\end{align*}

  \vspace{0.7cm}
  \item[2.4] \textit{Second cycle. CM-step for  constrained estimation of $\bLambda_g$ and $\bPsi_g$ :} \\
  Here our aim is to maximize $Q_2 \left( \btheta_2; \btheta^{(l)} \right)$ over $\btheta$, with $\btheta_1$ held fixed at $\btheta_1^{(l+1)}$. After some matrix algebra, this yields the updated ML-estimates
  \begin{align*}
\bLambda_g & = \bS^{(l+1)}_g {\bgamma^{(l)}}'_g [\bXi_g^{(l)}]^{-1} \\
\bPsi_g & =\mbox{diag}\left\{ \bS^{(l+1)}_g- \bLambda_g^{(l+1)} \bgamma_g^{(l)} \bS^{(l+1)}_g\right\} \,  
\end{align*}
where we denote by $\bS_g^{(l+1)}$ the sample scatter matrix in group $g$, for $g=1,\ldots,G$
\begin{equation*}
\bS_g^{(l+1)}  =(1/n_g^{(l+1)}) \sum_{i=1}^n z_{ig}^{(l+1)} \left( \bx_i - \bmu_g^{(l+1)} \right)  \left( \bx_i - \bmu_g^{(l+1)} \right)'. \\
\end{equation*}
    Along the iterations, due to the updates,
    it may happen that the $\bLambda_g$  matrices do not belong to the constrained parameter space $\bTheta_{c}$.
   In case that the eigenvalues of the $\bLambda_g\bLambda_g'$ matrices do not satisfy the required
    constraint (\ref{constraintOnLambda}), we obtain the ML solution in $\bTheta_{c}$ by projecting the unconstrained optimum into $\bTheta_{c}$. To this aim,
   the singular-value decomposition of  $\bLambda_g \bLambda'_g=\bL_g'\bE_g  \bL_g$ is considered,
    with $\bL_g$ being an orthogonal matrix and
    $\bE_g=\mbox{diag}(e_{g1},e_{g2},\allowbreak..., e_{gd})$ a diagonal matrix (notice that some of
    these $e_{gk}$ may be equal to 0 if $\bLambda_g \bLambda'_g$ is not full rank). 
    Truncated singular values are then defined as
$$
 [e_{gk}]_m =\min(c_{load}\cdot m, (\max(e_{gk},m))),  \mbox{ for } \,\,\,
 k=1,\ldots,d\mbox{ and }g=1,\ldots,G,
$$
   and $m$ being some threshold value. The loading matrices are finally updated
   as $\bLambda_g^{(l+1)}$ such that
   $\bLambda_g^{(l+1)} \big[\bLambda_g^{(l+1)}\big]' =\bL_g'\bE_g^*  \bL_g$
  with $$\bE_g^{*}=\text{diag}\left([e_{g1}]_{m_{\text{opt}}},[e_{g2}]_{m_{\text{opt}}},...,[e_{gd}]_{m_{\text{opt}}}
  \right)$$
   and $m_{\text{opt}}$ minimizing the real valued function
   \begin{equation} \label{m1}
   f_{load}(m) = \sum\limits_{g=1}^{G} \pi_g^{(l+1)} \sum\limits_{k=1}^{d}\left( \log
   \left( [e_{gk}]_m\right) +\frac{ e_{gk}}{[e_{gk}]_m}\right) .
   \end{equation}
  It may be mentioned here, in passing, that Proposition 3.2 in \cite{FriG13} shows that $m_{\text{opt}}$ can be obtained by
evaluating $2dG+1$ times the real valued function   $f_{load}(m)$ in
(\ref{m1}).

%
%

Given the $\bLambda_g^{(l+1)}$, we obtain the matrices
$$\bPsi_g  =\mbox{diag}\left\{ \bS^{(l+1)}_g-
\bLambda_g^{(l+1)} \bgamma_g^{(l)} \bS^{(l+1)}_g\right\}$$ 
which may not necessarily satisfy the required
constraint (\ref{constraintOnPsi}). In this
case, 
we set $$[\psi_{g,k}]_{m}=
\min(c_{noise} \cdot m,\, \max(\psi_{g,l},m)), \quad
\textrm{for} \,\,\, k=1,\ldots,d;\, g=1,\ldots,G,$$
and fix the optimal threshold value $m_{\text{opt}}$ by
minimizing the following real valued function
     \begin{equation} \label{mPsi}
   f_{noise}(m) \mapsto \sum\limits_{g=1}^{G} \pi_g^{(l+1)} \sum\limits_{k=1}^{p}\left( \log
   \left( [\psi_{g,k}]_m\right) +\frac{ \psi_{g,k}}{[\psi_{g,k}]_m}\right) .
   \end{equation}
As before, in  \cite{FriG13} it is shown that  $m_{\text{opt}}$  can
be obtained in a straightful way by   evaluating $2pG+1$ times
$f_{noise}(m)$ in  (\ref{mPsi}). Thus, $\bPsi_g^{(l+1)}$
is finally updated as
$$\bPsi_g^{(l+1)}=\mbox{diag}([\psi_{g,1}]_{m_{\text{opt}}},...,[\psi_{g,p}]_{m_{\text{opt}}}).$$
It is worth to remark that the given constrained estimation
provides, at each step, the parameters $\bPsi_g$ and $\bLambda_g$
maximizing the likelihood in the constrained parameter space
$\Theta_{c}$.
  \end{enumerate}

    \vspace{0.7cm}
  \item[3] \textit{Evaluate target function:}
  After applying the trimmed and constrained EM steps, and setting $z(\bx_i)=0$ if $i\in I$ and $z(\bx_i)=1$ if
$i\notin I$, the associated value of the target function (\ref{d1})
is evaluated. If convergence has not been achieved before reaching the maximum number of iterations $MaxIter$, results are discarded.  \end{enumerate}
    \vspace{0.7cm}
The set of parameters yielding
  the highest value of the target function (among the multiple runs) and the associated trimmed indicator function $z$ are returned as the  final output of the algorithm.  In the framework of model-based clustering, each unit is assigned to
one group, based on the maximum a posteriori probability. Notice, in passing, that we do not need a high number of initializations neither a high value for $MaxIter$, as we will see in Section \ref{sec:Numerical results}.

\section{Numerical studies}\label{sec:Numerical results}
In this section we present numerical studies, based on simulated and real data, to show the performance of the constrained and trimmed  AECM algorithm with respect to
unconstrained and/or untrimmed approaches.

\subsection{Artificial data}\label{sec:simdata}
We  consider here the following mixture of $G$
components of $d$-variate normal distributions.
To perform the estimation, we  consider 10 different random initial clusterings to initialize the algorithm at each run, as described in the previous section, and we retain the best solution.  The needed routines have been written in \texttt{R}-code \citep{R:2013}, and  are available from the authors upon request.

\paragraph{\textsc{Mixture: $G=3$, $d=6$, $q=2$, $N=150$.}} \ \\

\begin{figure} [h]  \begin{center}
        \includegraphics[width=8 cm, height=8 cm]{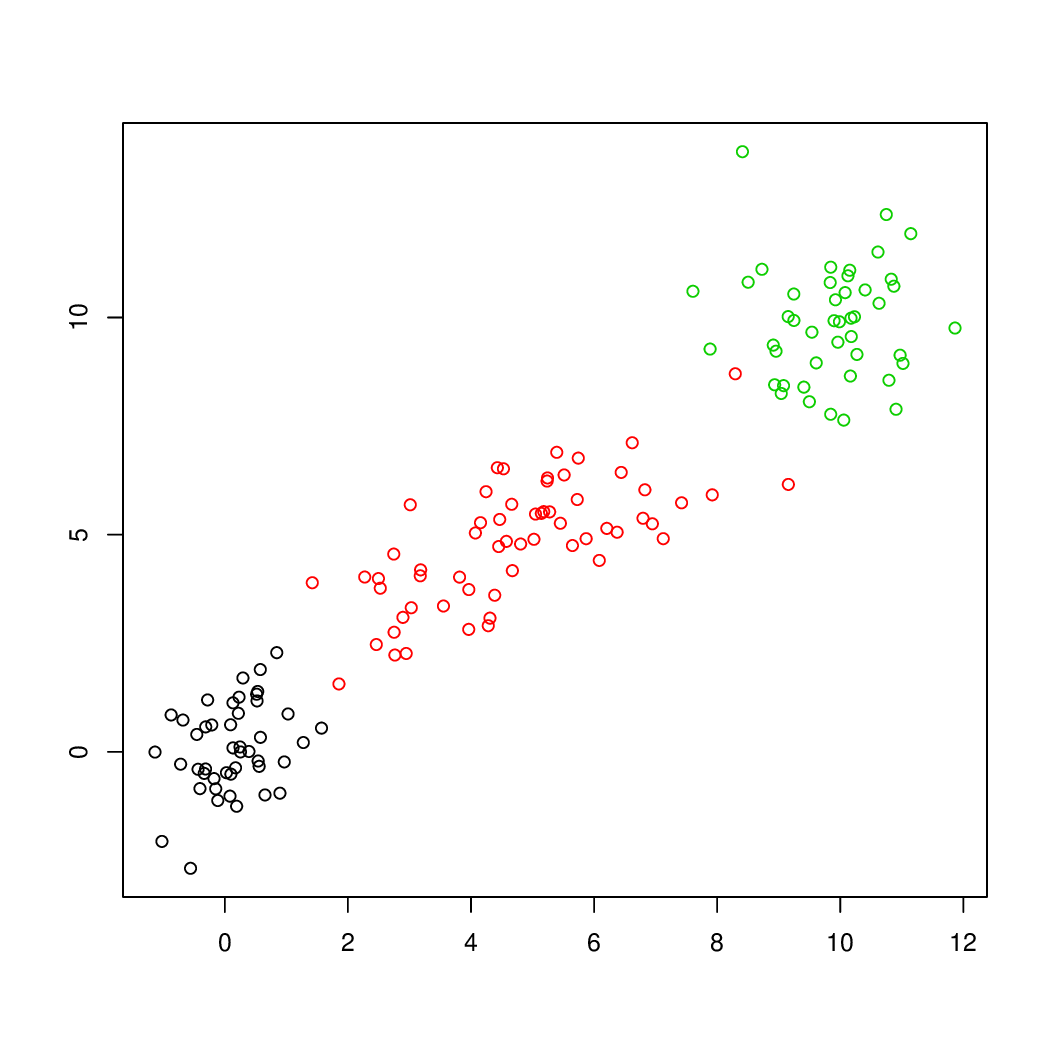}
            \end{center}
    \caption{\footnotesize A specimen of 150 data points generated from the mixture  (the first two coordinates are plotted, groups in black, red and green)
    }
    \label{fig:Mixdata}
\end{figure}

The sample has been generated with weights $\bpi = (0.3, 0.4, 0.3)'$ according to the following parameters:
\begin{footnotesize}
  \vspace{-3mm}\\
   \begin{align*}
    \renewcommand{\arraystretch}{0.75}
 \bmu_1 &= (0,0,0,0,0,0)'  &  \bPsi_1 &= \mbox{diag}(0.1,0.1,0.1,0.1,0.1,0.1) \\
 \bmu_2 &= (5,5,5,5,5,5)'  & \bPsi_2 &= \mbox{diag}(0.4,0.4,0.4,0.4,0.4,0.4) \\
 \bmu_3 &= (10,10,10,10,10,10)'  & \bPsi_3 &= \mbox{diag}(0.2,0.2,0.2,0.2,0.2,0.2)
  \end{align*}
  \vspace{-8mm}\\
 \begin{gather*}
 \renewcommand{\arraystretch}{0.75}
\bLambda_1 = \begin{pmatrix} 0.50 & 1.00 \\ 1.00 & 0.45 \\  0.05 & -0.50 \\ -0.60 & 0.50 \\ 0.50 & 0.10 \\ 1.00 & -0.15 \end{pmatrix}  \quad  \quad
\bLambda_2 = \begin{pmatrix} 0.10 & 0.20 \\ 0.20 & 0.50 \\  1.00 & -1.00 \\ -0.20 & 0.50 \\ 1.00 & 0.70 \\ 1.20 & -0.30 \end{pmatrix} \quad  \quad
 \bLambda_3 = \begin{pmatrix} 0.10 & 0.20 \\ 0.20 & 0.00 \\  1.00 & 0.00 \\ -0.20 & 0.00 \\ 1.00 & 0.00 \\ 0.00 & -1.30 \end{pmatrix}.\\
\end{gather*}
  \vspace{-8mm}\\
  \end{footnotesize}
Figure \ref{fig:Mixdata} shows a specimen of randomly generated data from the given mixture.

Our analysis begins by running the AECM algorithm on the generated sample,  and considering the following six settings,
namely:
\begin{footnotesize}
\begin{enumerate}
\item[S1.] a "virtually" unconstrained approach (i.e. $c_{\rm noise}=c_{\rm load}=10^{10}$) without trimming ($\alpha=0$),
\item[S2.] an adequate constraint on $\bPsi_g$, no constraint on $\bLambda_g$ ($c_{\rm noise}=5, c_{\rm load}=10^{10}$) and no trimming ($\alpha=0$),
\item[S3.] adequate constraints on $\bPsi_g$ and $\bLambda_g$ ($c_{\rm noise}=5$, $c_{\rm load}=3$), and still no trimming ($\alpha=0$),
\item[S4.] a "virtually" unconstrained approach (i.e. $c_{\rm noise}=c_{\rm load}=10^{10}$) with trimming ($\alpha=0.06$),
\item[S5.] an adequate constraint on $\bPsi_g$, no constraint on $\bLambda_g$ ($c_{\rm noise}=5, c_{\rm load}=10^{10}$), with trimming ($\alpha=0.06$),
\item[S6.] adequate constraints on $\bPsi_g$ and $\bLambda_g$  ($c_{\rm noise}=5$, $c_{\rm load}=3$), with trimming ($\alpha=0.06$)
\end{enumerate}
  \end{footnotesize}
It is worth noticing that when setting $c_{noise}=10^{10}$ we want to discard singularities, and allow the estimation to move in a wide parameter space that contains the global maximum, among several local ones. In this situation the estimation could incur into spurious solutions. We expect that the algorithm improve its performances when giving the "right" constraints. The adequate constraints can by evaluated by obtaining the maximum ratio among the eigenvalues of $\bPsi_g$ and among the singular values of $\bLambda_g$.
We have that the singular values of $\bLambda_1$ are $( 3.069, 1.528)$, of
$\bLambda_2$ are (3.777, 1.873) and of $\bLambda_3$ are (2.091, 1.729) hence we derive $c_{load} \geq 2.471$;
while the diagonal elements of $\bPsi_g$ are 0.1, 0.4, and 0.2 so that $c_{noise} \geq 4$.
We applied also trimming to the artificially generated data, to see the effect of an unneeded elimination of the outermost points in the model estimation and subsequent classification. We evaluate the performance of the algorithm by calculating the average misclassification error $\eta$, over $100$ repetitions of the estimation procedure. The misclassification error is defined as the relative frequency of points of the sample erroneously labeled, taking into account that noise and pointwise contamination (when added) should be identified, as they virtually do not belong to the three groups. We see that the algorithm, applied without trimming, give a perfect classification with and without constraints, due to fact that estimation is performed along 10 random initializations. While adding trimming,  the misclassification error is almost equal to  the trimming level (as expected).  The results are summarized in Table \ref{tab:misclass_error}. Moreover, we observed that the other parameters, such as the means $\bmu_g$, and $\bPsi_g$, $\bLambda_g$ for $g=1,2,3$ are close to the values from which the data have been generated.

\medskip
  \begin{tiny}
\begin{table}[ht]
\footnotesize
\centering
\caption{\footnotesize Misclassification error $\eta$ (average on 100 repetitions of the estimation procedure) of the AECM algorithm with settings S1-S6, applied on the artificially generated data \newline}
\label{tab:misclass_error}       
\begin{tabular}{rcccccc}
  \hline\hline
  & S1 & S2 & S3 & S4 & S5 & S6 \\ \cline{2-7}
$c_{\rm noise}$ &  $10^{10}$ & 5 &  5 &   $10^{10}$ & 5 &  5  \\
$c_{\rm load}$ &  $10^{10}$    &  $10^{10}$   &  3 &  $10^{10}$    & $10^{10}$   &  3  \\
 $\alpha$ & 0 &  0&  0 &  0.06& 0.06&  0.06 \\
  \hline 
$\eta $ & 0.33\% & 0.04\% & 0.00\% & 6.45\% & 6.13\% & 6.00\% \\
 \hline\hline
\end{tabular}
\end{table}
  \end{tiny}
\medskip

Afterwards, we have considered 3 different scenarios.
\begin{enumerate}[topsep=1.5pt,itemsep=-1.8ex,partopsep=1.5ex,parsep=1.5ex]
 \item [D+N:] 10 points of uniform noise have been added around
 the data,  
 \item[D+PC:]  10 points of pointwise contamination have been added outside the range of
the data,
\item[D+N+PC:]  both the uniform noise and the pointwise contamination have been added to the data.
\end{enumerate}
We applied the algorithm to the different datasets in the six previous settings S1-S6 (i.e. with/without constraints and trimming) and we calculated the misclassification error.  Results in the first row of Table \ref{tab:2} show that trimming is very effective to identify and discard noise in the data, and constraints contribute slightly, to reach  perfect classification.
The misclassification error (reported in the second row of Table \ref{tab:2}) shows that we need trimming \textit{and} constraints to achieve a very good behavior of the algorithm. Noise and pointwise contamination could cause very messy estimation, as it is seen in the first three columns of the Table, where we only rely/do not rely on constraints. Further, we observe that trimming is a good strategy when dealing with uniform noise, but it is not able to resist pointwise contamination. If we want to be protected against all type of data contamination we do need both the use of constrained estimation and trimming.

\medskip
\begin{table}[h]
\footnotesize
\centering
\caption{\footnotesize Misclassification error $\eta$ (average on 100 repetitions of the estimation procedure)  of the AECM algorithm   with settings S1-S6, applied on different data sets \newline}
\label{tab:2}       

\begin{tabular}{rcccccc}
  \hline\hline
   & S1 & S2 & S3 & S4 & S5 & S6 \\ \cline{2-7}
  $c_{\rm noise}$ & $10^{10}$  & 5 &  5 &   $10^{10}$ & 5 &  5  \\
$c_{\rm load}$ & $10^{10}$    &  $10^{10}$    &  3 &  $10^{10}$   & $10^{10}$    &  3  \\
 $\alpha$ & 0 &  0&  0 &  0.06& 0.06&  0.06 \\
   \hline 
D+N  & 0.3348 & 0.4856 & 0.5357 & 0.0305 & 0.0033 & 0.0000 \\
D+PC & 0.2811 & 0.2659 & 0.2837 & 0.0465 & 0.0071 & 0.0031 \\
D+N+PC & 0.4035 & 0.5299 & 0.5294 & 0.0918 & 0.0124 & 0.0064 \\
\hline\hline
\end{tabular}
\end{table}
\medskip


\subsubsection{Properties of the estimators for the mixture parameters}

Now, we want to perform a second analysis on this artificial data
and our main interest here is in assessing the effect of trimming
and constraints on the properties of the model estimators. Namely,
we will estimate their bias and mean square error  when the data is
affected by noise and/or pointwise contamination.
 We will consider the same four scenarios we considered before, i.e.:
\begin{enumerate}[topsep=1.5pt,itemsep=-1.8ex,partopsep=1.5ex,parsep=1.5ex]
\item[D:] the artificially generated data,
\item[D+N:] the data  with added noise,
\item[D+PC:] the data with added pointwise contamination,
\item[D+N+PC:] the data with added noise and pointwise contamination.
\end{enumerate}
We apply in the four scenarios the algorithm for estimating a trimmed MFA model, exploring the six settings on $c_{noise}$,
$c_{load}$ and
 $\alpha$ that have been shown in Table \ref{tab:2}. For sake of space, we report our results only on the  more interesting
 cases. 

The benchmark of all simulations is given by the results that we
obtain on artificial data  drawn from a given MFA without
outliers,
and they are shown in the first column of Table
\ref{tab:summaryFIRSTresults}. In each experiment, we draw 1000 times a
sample of size $n=150$ from the mixture described at the beginning
of this Section, and we estimate the model parameters for the
trimmed MFA using the algorithm presented in the previous Section
\ref{se3_2}. We set $c_{noise}=c_{load}=10^{10}$ and $\alpha=0$ for
this first case, as no outliers are added to the samples.

Notice that the considered estimators in each component are vectors (apart from $\pi_g$ which are scalar quantities, for $g=1,\ldots,G$). We are interested in providing  synthetic measures of their properties, such as bias and mean square error (MSE). As usual, let $\hat{T}$ be an estimator for the scalar parameter $t$, then the bias of $\hat{T}$ is  given by $bias(\hat{T})=\EE(\hat{T})-t$, i.e. it is the signed absolute deviation of the expected value $\EE(\hat{T})$  from  $t$. 
Therefore, we would have 6 biases for each component of the mean
$\mu_g$, 6 for diag($\Psi_g$) and 12  for $\Lambda_{g}$. On the
other side, MSE is defined as a scalar quantity, namely $
\EE(|\hat{T}-t|^2)=\textrm{trace}(Var(\hat{T}))+\textrm{bias}(\hat{T})^2$,
also for vector estimators. Hence, we  adopted a  synthesis of each
parameter biases by considering the mean of their absolute values on
each component. Below the bias, in Table \ref{tab:summaryFIRSTresults},
we provide the MSE in parenthesis.

Then, the second experiment consists in drawing 1000 samples as
before, and adding 10 points of random uniform noise to each of
them; the bias and mean square error for the model estimators
increase dramatically, with  $c_{noise}=c_{load}=10^{10}$ and
$\alpha=0$ (results are  displayed in second column of Table \ref{tab:summaryFIRSTresults}). On
the other hand, results go back to the same order of magnitude as
the benchmark  if we impose $c_{noise}=5, c_{load}=3$ and
$\alpha=0.06$,   as it is shown in the second column of Table
\ref{tab:summarySECONDresults}.

The third experiment is based on 1000 samples, with 10 points
of  pointwise contamination randomly added. We observed a huge
increase of the bias and mean square error for the model estimators,
without appropriate constraints and level of trimming (see results in third column of Table \ref{tab:summaryFIRSTresults}), but whenever
we run the algorithm with $c_{noise}=5, c_{load}=3$ and $\alpha=0.06$
we came back to results very close to the benchmark, shown in the third
column of Table \ref{tab:summarySECONDresults}.

The fourth experiment  has been developed by considering added random noise and pointwise contamination to the 1000 drawn samples. The results on bias and mean square error for the case of estimating the trimmed MFA    with  $c_{noise}=c_{load}=10^{10}$ and $\alpha=0$,  in the fourth column of Table \ref{tab:summaryFIRSTresults}, show the harmful effects of distorted inference. On the other side, when we applied reasonable constraints $c_{noise}=5$, $ c_{load}=3$ and a trimming level $\alpha=0.12$ to cope with the added outliers, we got the results in the fourth column of  Table  \ref{tab:summarySECONDresults}.  We see that robust inference allows reduced bias and mean square error, even in case of both sparse outliers and concentrated leverage points.

Finally the  scheme of simulations on the four data sets, in the six estimation settings, have been repeated considering a triple sample
size ($n=450$). All results are summarized in  Table \ref{tab:summaryFIRSTresults}  when $c_{noise}=c_{load}=10^{10}$ and $\alpha=0$ to be compared with results in Table \ref{tab:summarySECONDresults}  in which appropriate constraints and trimming have been used along the estimation, to see the improved properties of the estimators when the sample size increases.

\begin{table}[ht]
\centering
\footnotesize{\caption{ \footnotesize Case without trimming and constraints.  \newline \footnotesize Bias as 
the sum of absolute deviations, 
followed by  MSE  (in parentheses) of the parameter estimators $\hat{\pi}_{i},  \hat{\bmu}_{i} ,
\hat{\bPsi}_{i} , \hat{\bLambda}_{i}$, for $i=1,2,3$ when dealing
with  different datasets. The column labels refer to the different
scenarios, namely D stays for \textit{data}, D+N stays for
\textit{data and added noise}, D+PC stays for \textit{data and
pointwise contamination}, D+N+PC stays for \textit{data with added
noise and pointwise contamination.\newline} } \label{tab:summaryFIRSTresults}
\renewcommand{\arraystretch}{0.5}
\begin{tabular}{lccccc}
  \hline\hline\\
  &	D 	&	D+N	&D+PC	&	D+N+PC	&3*(D+N+PC) \\ 
  \\
  \hline
  \\
$\hat{\tau}_1$ 
   & 0.0013 & 0.124 & 0.1219 & 0.1893 & 0.1649  \\ 
   & (0) & (0.0166) & (0.0153) & (0.038) & (0.0312) \\ 
   \\
  $\hat{\tau}_2$ 
   & 0.0065 & 0.0877 & 0.1359 & 0.324 & 0.2097 \\ 
   & (0) & (0.0089) & (0.0189) & (0.1072) & (0.048) \\ 
   \\
  $\hat{\tau}_3$ 
   & 0.0053 & 0.2118 & 0.2579 & 0.1347 & 0.0448 \\ 
   & (0) & (0.0461) & (0.067) & (0.0204) & (0.006) \\ 
   \\
  $\hat{\bmu}_1$ 
   & 0.018 & 1.506 & 1.955 & 11.534 & 13.093 \\ 
   & (0.305) & (31.736) & (39.371) & (659.863) & (1140.549) \\ 
   \\
    $\hat{\bmu}_2$ 
  & 0.006 & 5.15 & 5.87 & 1.845 & 3.728  \\ 
   & (0.497) & (345.478) & (133.962) & (99.898) & (207.869)  \\ 
   \\
  $\hat{\bmu}_3$ 
   & 0.059 & 11.651 & 2.63 & 11.949 & 8.159  \\ 
   & (0.712) & (729.452) & (17.135) & (867.905) & (790.799)   \\ 
   \\
    $\hat{\bPsi}_{1}$ 
   & 0.013 & 0.435 & 0.043 & 0.564 & 1.274  \\ 
   & (0.027) & (11.373) & (0.054) & (115.608) & (272.187)   \\ 
   \\
  $\hat{\bPsi}_{2}$ 
   & 0.066 & 4.622 & 0.203 & 3.919 & 9.339  \\ 
   & (0.172) & (1520.318) & (0.79) & (651.607) & (1819.385) \\ 
   \\
  $\hat{\bPsi}_{3}$ 
   & 0.239 & 15.986 & 0.38 & 14.736 & 25.429 \\ 
   & (2.039) & (5634.188) & (1.891) & (4557.463) & (10196.275) \\ 
   \\
  $\hat{\bLambda}_{1}$ 
   & 0.534 & 0.534 & 0.534 & 0.498 & 0.522 \\ 
   & (82.817) & (82.817) & (96.75) & (300.57) & (361.565) \\ 
   \\
  $\hat{\bLambda}_{2}$ 
   & 0.608 & 0.608 & 0.551 & 0.642 & 0.653  \\ 
  & (172.601) & (172.601) & (86.747) & (304.718) & (633.379) \\ 
  \\
  $\hat{\bLambda}_{3}$ 
   & 0.335 & 0.335 & 0.354 & 0.373 & 0.341 \\ 
   & (404.326) & (404.326) & (53.063) & (284.401) & (326.672) \\ 
   \\
   \hline\hline
\end{tabular}
}
\end{table}
\medskip

\begin{table}[h!]
\centering
\footnotesize{\caption{\footnotesize Case with trimming and constraints.   \newline
\footnotesize Bias as 
the sum of absolute deviations, 
followed by  MSE  (in parentheses) of the parameter estimators $\hat{\pi}_{i},  \hat{\bmu}_{i} ,
\hat{\bPsi}_{i} , \hat{\bLambda}_{i}$, for $i=1,2,3$ when dealing
with  different datasets. The column labels refer to the different
scenarios, namely D stays for \textit{data}, D+N stays for
\textit{data and added noise}, D+PC stays for \textit{data and
pointwise contamination}, D+N+PC stays for \textit{data with added
noise and pointwise contamination. \newline} } \label{tab:summarySECONDresults}
\renewcommand{\arraystretch}{0.5}
\begin{tabular}{lccccc}
 \hline\hline
  \\
  &	D 	&	D+N	&D+PC	&	D+N+PC	&3*(D+N+PC) \\
  \\ \hline
  \\
$\hat{\tau}_1$ 
   & 
   0.0151 & 0.0002 & 0.0011 & 0 & 0.0006 \\ 
   & 
   (0.0002) & (0) & (0) & (0) & (0) \\ 
 \\
  $\hat{\tau}_2$ 
   & 
   0.0195 & 0.0014 & 0.0011 & 0.0034 & 0.0006\\ 
   & 
   (0.0004) & (0) & (0) & (0) & (0)\\ 
 \\
  $\hat{\tau}_3$ 
   & 
   0.0044 & 0.0012 & 0.0001 & 0.0034 & 0.0001\\ 
   & 
   (0) & (0) & (0) & (0) & (0)\\ 
 \\
  $\hat{\bmu}_1$ 
   & 
   0.006 & 0.010 & 0.017 & 0.006 & 0.001\\ 
   & 
   (0.117) & (0.159) & (0.215) & (0.226) & (0.038) \\ 
 \\
  $\hat{\bmu}_2$ 
  & 
  0.006 & 0.004 & 0.002 & 0.007 & 0.009 \\ 
   & 
   (0.190) & (0.219) & (0.165) & (0.703) & (0.158)\\ 
 \\
  $\hat{\bmu}_3$ 
   & 
    0.008 & 0.020 & 0.004 & 0.062 & 0.018 \\ 
   & 
   (0.177) & (0.302) & (0.154) & (0.787) & (0.231)\\ 
   \\
    $\hat{\bPsi}_{1}$ 
   & 
   0.013 & 0.029 & 0.026 & 0.032 & 0.022\\ 
   & 
   (0.010) & (0.025) & (0.024) & (0.035) & (0.009) \\ 
   \\
  $\hat{\bPsi}_{2}$ 
   & 
   0.066 & 0.044 & 0.045 & 0.046 & 0.042 \\ 
   & 
   (0.089) & (0.082) & (0.081) & (0.102) & (0.058) \\ 
    \\
  $\hat{\bPsi}_{3}$ 
   & 
   0.066 & 0.072 & 0.075 & 0.076 & 0.075\\ 
   & 
   (0.108) & (0.158) & (0.154) & (0.178) & (0.158) \\ 
   \\
  $\hat{\bLambda}_{1}$ 
   & 
   0.516 & 0.512 & 0.516 & 0.546 & 0.524\\ 
   & 
   (13.168) & (14.828) & (12.992) & (14.711) & (13.674)\\ 
   \\
  $\hat{\bLambda}_{2}$ 
   & 
   0.568 & 0.569 & 0.569 & 0.586 & 0.578 \\ 
  & 
  (11.049) & (12.311) & (12.036) & (13.875) & (12.832)\\ 
 \\
  $\hat{\bLambda}_{3}$ 
   & 
   0.330 & 0.353 & 0.354 & 0.341 & 0.342 \\ 
   & 
   (9.377) & (11.164) & (12.259) & (13.069) & (13.064)\\ 
   \\
   \hline\hline
\end{tabular}
}
\end{table}
\medskip
\normalsize

\begin{figure}
    \begin{center}
    \includegraphics[width=7.9 cm ]{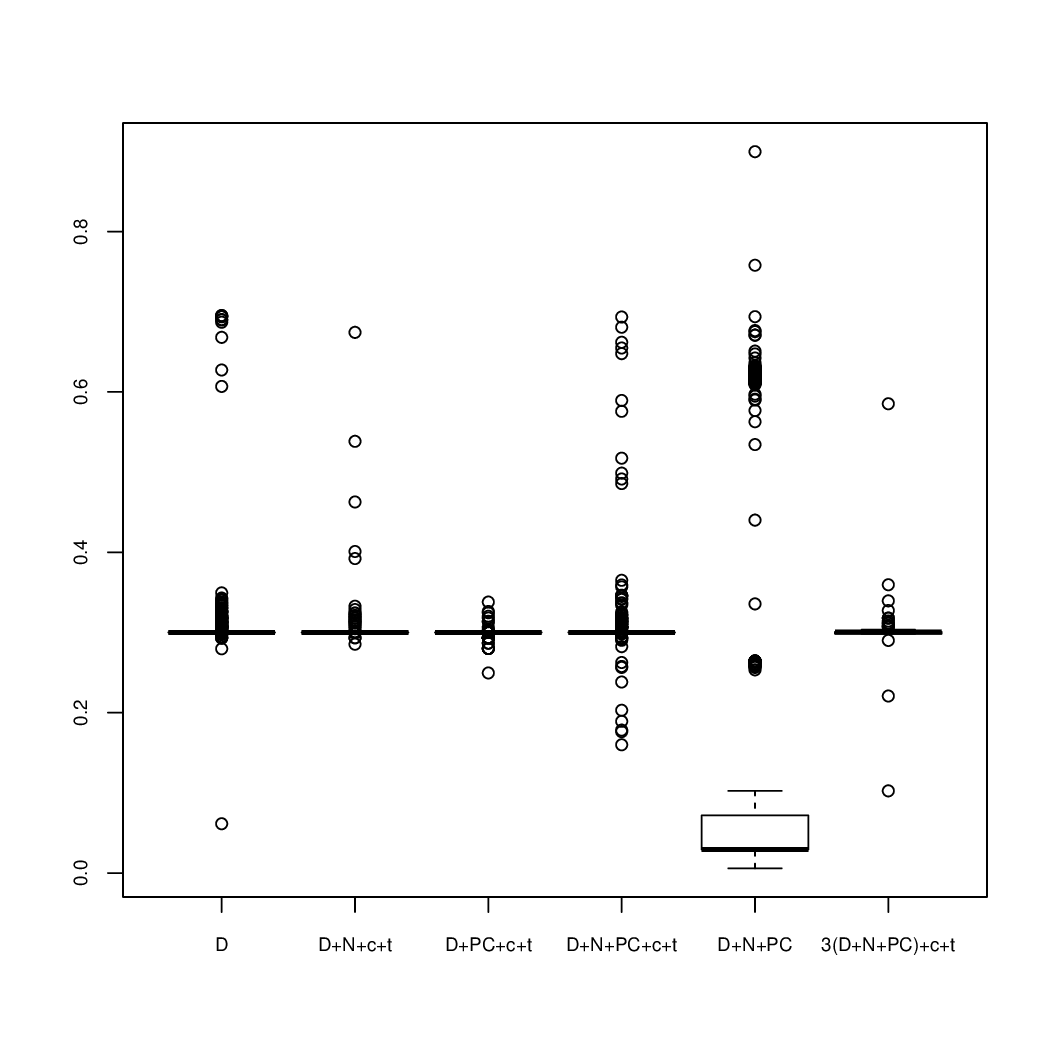}
    \includegraphics[width=7.9 cm]{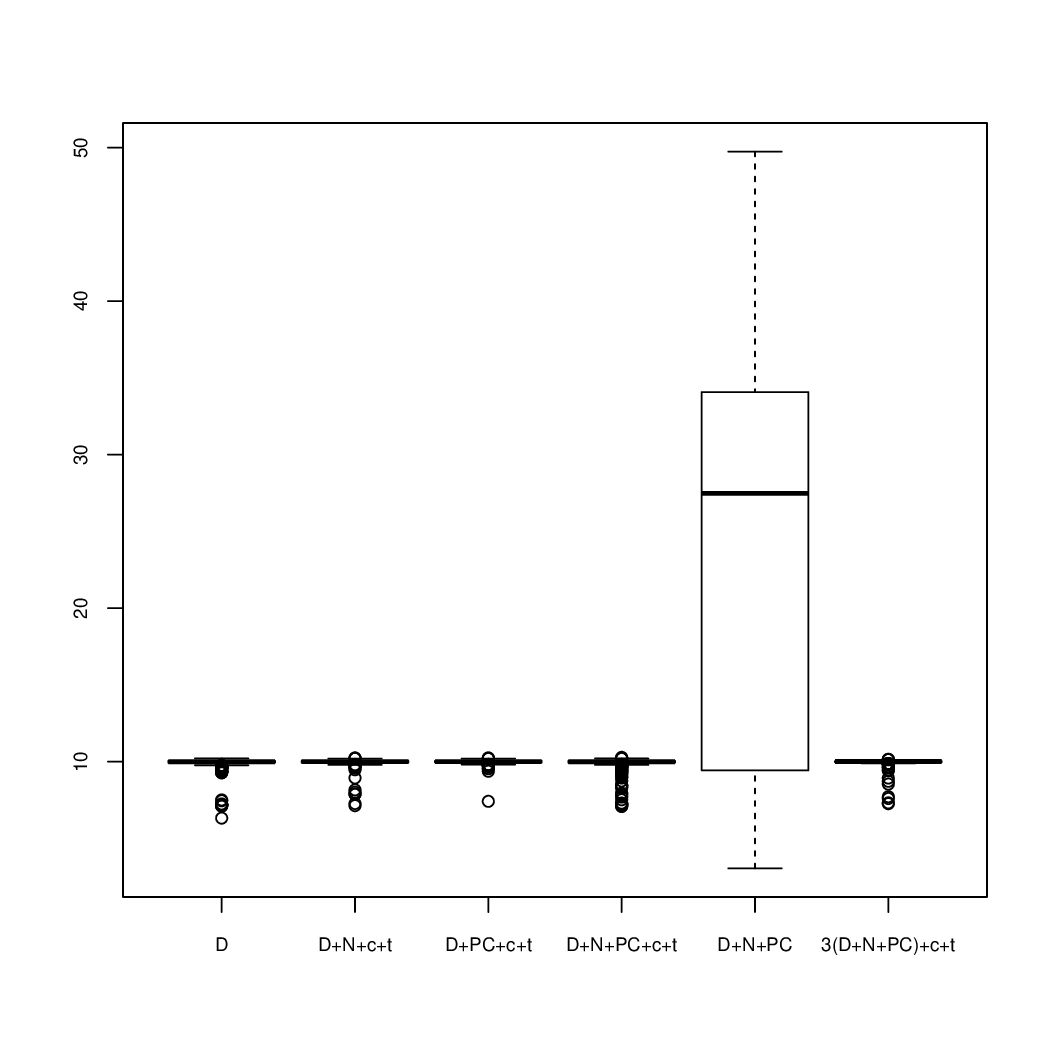}\\
    \includegraphics[width=7.9 cm]{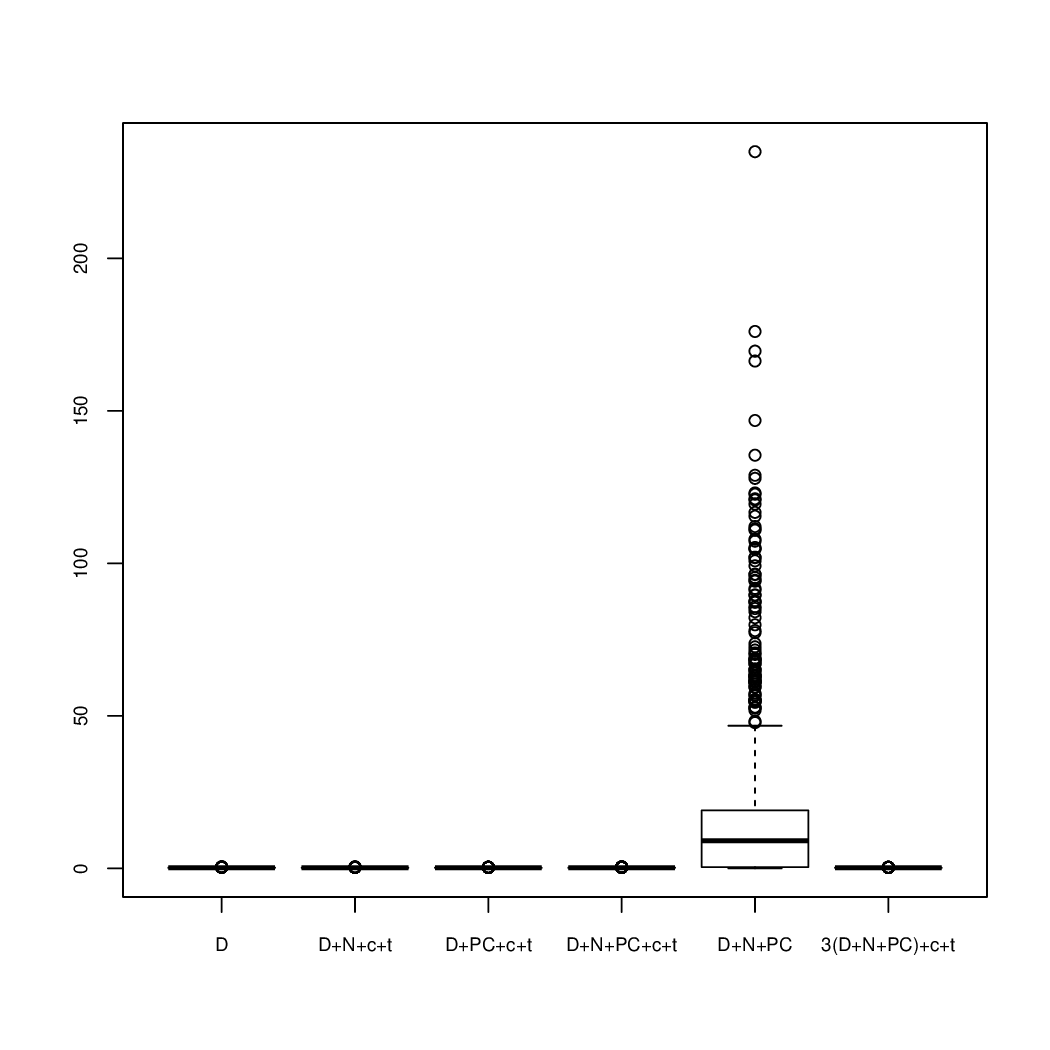}
    \includegraphics[width=7.9 cm]{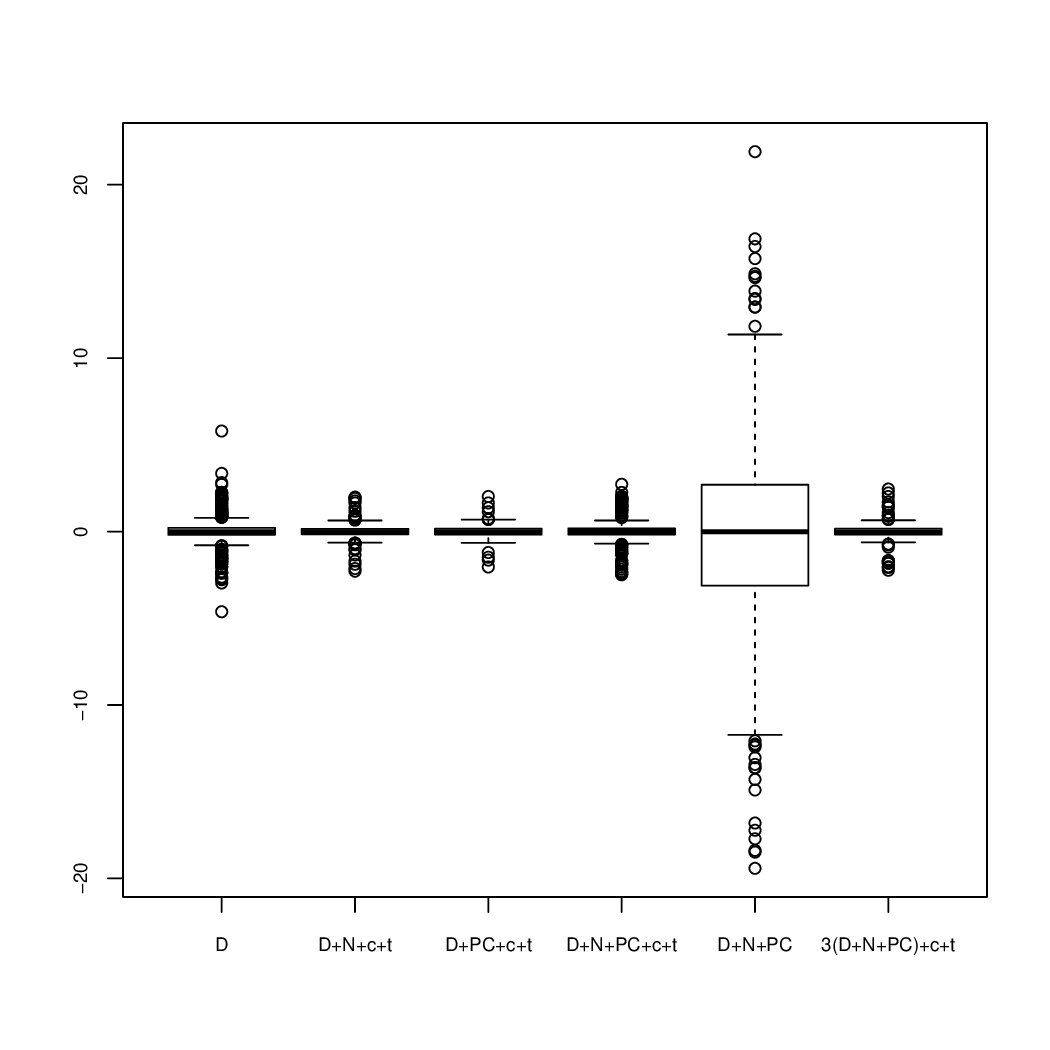}
    \end{center}
    \caption{\footnotesize Boxplots of some of the simulated distributions of the mixture parameters: \newline $\qquad \qquad $ the simulated distribution of $\hat{\pi}_3$, estimator for  $\pi_3=0.3$ (upper panel); \newline the simulated distribution of $\hat{\mu}_3[1]$, estimator for $\mu_3[1]=10$ (second panel from above);  \newline the simulated distribution of $\hat{\Psi}_3[1,1]$, estimator for $\Psi_3[1,1]=0.2$ (third panel from above); and  \newline the simulated distribution of $\hat{\Lambda}_3[1,1]$, estimator for $\Lambda_3[1,1]=0.1$ (fourth panel). \newline The labels on the horizontal axis refers to the six settings, namely ``D" stays for \textit{data}, ``D+N" stays for \textit{data and added noise}, ``D+N+PC" stays for \textit{data with added noise and pointwise contamination}, while ``c + t" has been added to denote the cases in which the estimation has been performed  \textit{using constraints and trimming}.}\label{fig:boxplots3}
\end{figure}

The distributions of the estimators for the model parameters can be represented through some box plots, and some of them are shown in Figure \ref{fig:boxplots3}, namely with reference to $\hat{\pi_3}$ (upper panel), $\hat{\mu}_3[1,1]$ (second panel),  $\hat{\Psi}_3[1,1]$ (third panel) and $\hat{\Lambda}_3[1,1]$ (bottom panel). We can see, in a direct comparison, that the estimation algorithm with adequate trimming and constraints is able to resist all type of outlying data. In each panel, the first boxplot on the left provides the benchmark of the following five ones, as it shows the distribution of the estimator when the data has been drawn from the mixture. The second boxplot (from  left to right) in each panel shows the  distribution of the estimator when we employ constraints and trimming along the estimation on data and added uniform noise. The third boxplot refers to the case in which we deal with data and pointwise contamination, and the good results are obtained because we are employing constraints and trimming. The fourth box plot has been obtained when considering both noise and pointwise contamination, and robust estimation. The fifth box plot shows the effects of noise and pointwise contamination when the estimation procedure does not employ constraints and trimming. Finally the sixth box plot in all panels reports the case of robust estimation performed on a triple sample size, still with noise and pointwise contamination.

\scriptsize
\normalsize

\subsection{Real data: the AIS data set}\label{sec:realdata}
\begin{figure} [h!]  \begin{center}
        \includegraphics[width=9 cm, height=9 cm]{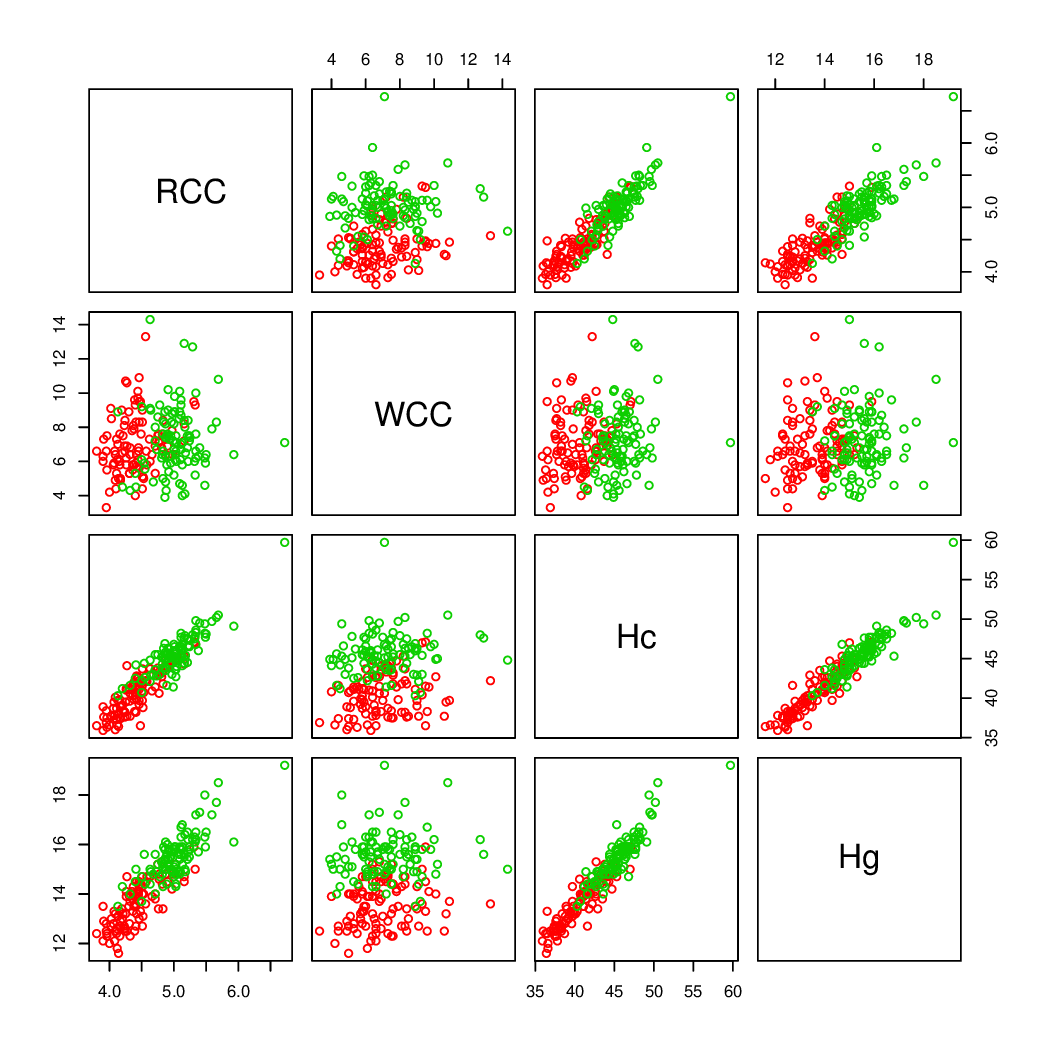}
            \end{center}
    \caption{\footnotesize Scatterplot of some pairs of the AIS variables (female data in red, male in blue)}\label{fig:AISdata}
\end{figure}

As an illustration, we apply the proposed technique to the Australian Institute of Sports (AIS) data, which is a famous benchmark dataset in the multivariate literature, originally reported by \citet{cook2009introduction}  and subsequently analyzed by \citet{azzalini1996multivariate}, among many other authors.
The dataset consists of $p=11$ physical and hematological measurements on 202 athletes (100 females and 102 males) in different sports, and is available within the \textrm{R} package \textit{sn} (Azzalini, 2011). The observed variables are: red cell count (RCC), white cell count (WCC), Hematocrit (Hc), Hemoglobin (Hg), plasma ferritin concentration (Fe), body mass index, weight/height$^2$ (BMI), sum of skin folds (SSF), body fat percentage (Bfat), lean body mass (LBM), height, cm (Ht), weight, kg (Wt), a part from Sex and kind of Sport.   A partial scatterplot of the AIS dataset is given in Figure \ref{fig:AISdata}, and Table \ref{tab:sumAIS} provides summary information.
\medskip
\begin{table}[ht]
\centering
\footnotesize{\caption{\footnotesize Summary information for the AIS dataset \newline} \label{tab:sumAIS} 
\begin{tabular}{lrrrrrr|rrrrrr}
\\
  \hline  \hline
  \footnotesize
&\multicolumn{6}{c}{female athletes}&\multicolumn{6}{c}{male athletes}\\
 & \cline{1-12}
 & min & Q1 & Me  & Q3 & max & mean & min & Q1 & Me  & Q3 & max & mean \\
  \hline
$RCC$ & 3.8 & 4.2 & 4.4 & 4.5 & 5.3 & 4.4 & 4.1 & 4.9 & 5.0 & 5.2 & 6.7 & 5.0 \\
  $WCC$ & 3.3 & 5.8 & 6.7 & 8.0 & 13.3 & 7.0 & 3.9 & 6.0 & 7.1 & 8.4 & 14.3 & 7.2 \\
  $Hc$ & 35.9 & 38.3 & 40.6 & 42.3 & 47.1 & 40.5 & 40.3 & 44.2 & 45.5 & 46.8 & 59.7 & 45.6 \\
  $Hg$ & 11.6 & 12.7 & 13.5 & 14.3 & 15.9 & 13.6 & 13.5 & 14.9 & 15.5 & 15.9 & 19.2 & 15.6 \\
  $Fe$ & 12.0 & 36.0 & 50.0 & 71.5 & 182.0 & 57.0 & 8.0 & 55.0 & 89.5 & 123.5 & 234.0 & 96.4 \\
  $BMI$ & 16.8 & 20.3 & 21.8 & 23.4 & 31.9 & 22.0 & 19.6 & 22.3 & 23.6 & 25.2 & 34.4 & 23.9 \\
  $SSF$ & 33.8 & 59.3 & 81.8 & 107.4 & 200.8 & 87.0 & 28.0 & 37.5 & 47.7 & 58.1 & 113.5 & 51.4 \\
  $Bfat$ & 8.1 & 13.2 & 17.9 & 21.4 & 35.5 & 17.8 & 5.6 & 7.0 & 8.6 & 10.0 & 19.9 & 9.3 \\
  $LBM$ & 34.4 & 51.9 & 54.9 & 59.4 & 73.0 & 54.9 & 48.0 & 68.0 & 74.5 & 80.8 & 106.0 & 74.7 \\
  $Ht$ & 148.9 & 171.0 & 175.0 & 179.7 & 195.9 & 174.6 & 165.3 & 179.7 & 185.6 & 191.0 & 209.4 & 185.5 \\
 $ Wt$ & 37.8 & 60.1 & 68.1 & 74.4 & 96.3 & 67.3 & 53.8 & 73.9 & 83.0 & 90.3 & 123.2 & 82.5 \\
   \hline  \hline
\end{tabular}
}
\end{table}
\medskip

\begin{figure}
\begin{center}
     \includegraphics[height=8 cm, width=7.8 cm]{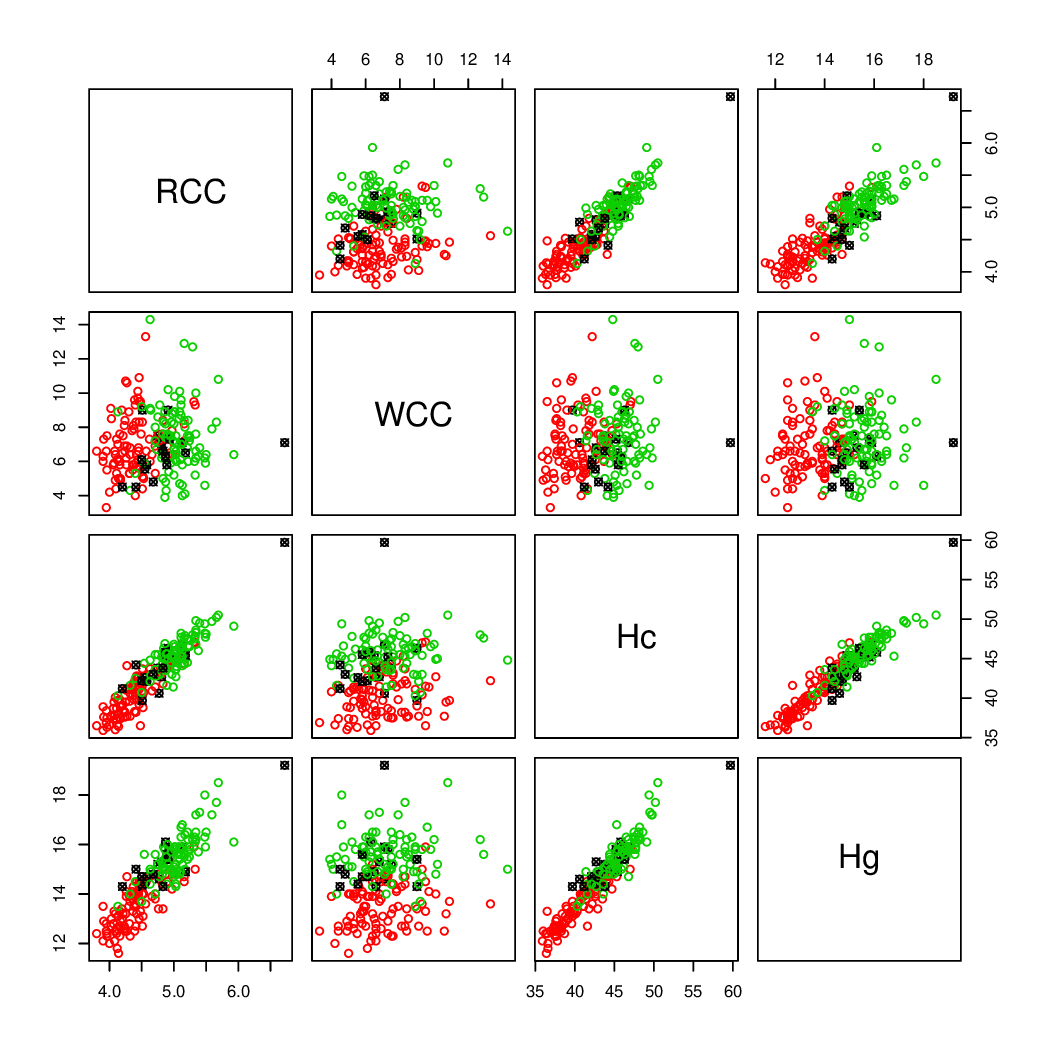}
       \includegraphics[width=7.8 cm, height=8.2 cm]{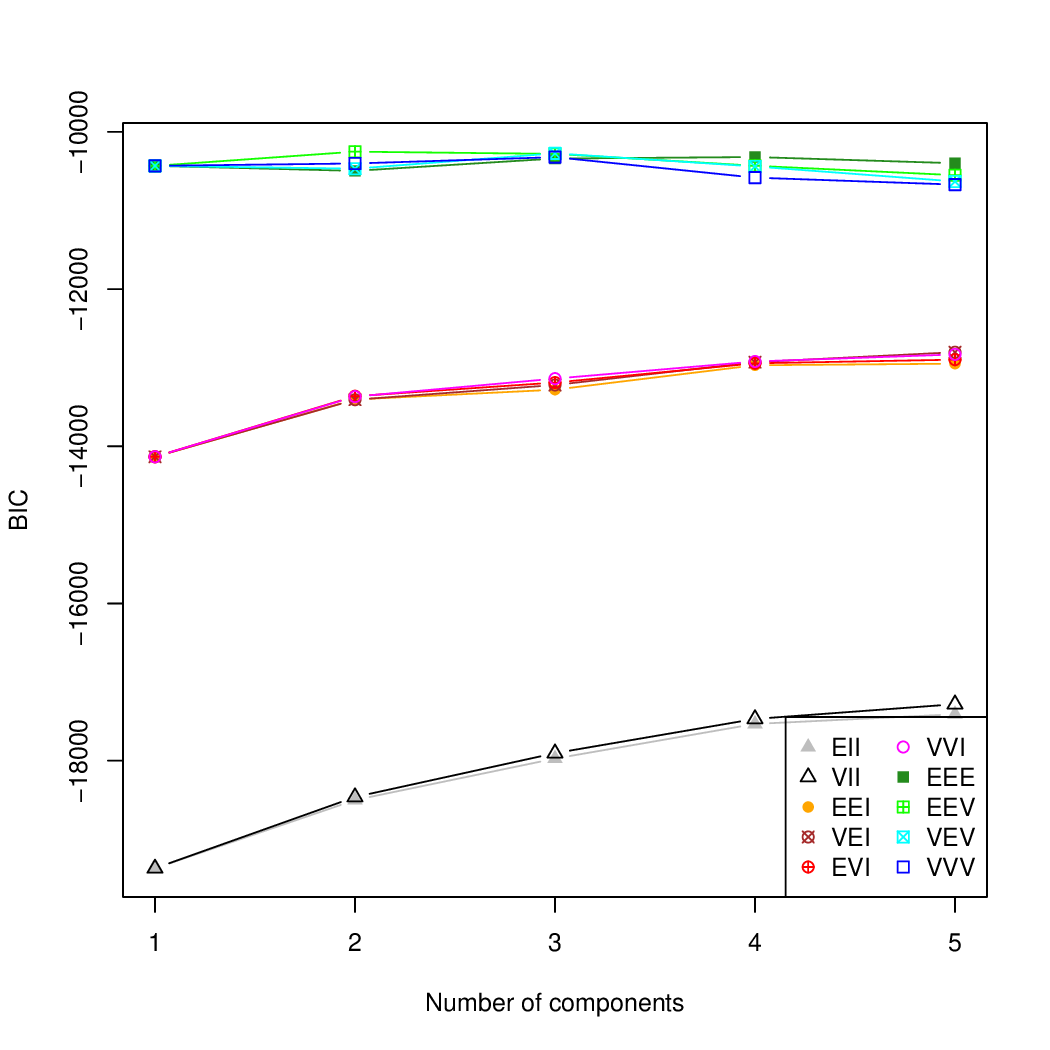}
\end{center}
\caption{\footnotesize The classification of AIS data obtained through the best model from Mclust (left panel, with female data in red, male in blue, misclassified units as black circlecrosses), and the graphical tool for model selection (right panel)}\label{fig:MisclAISdata}
\end{figure}

Our purpose is to provide a  model for the entire dataset, and classify athletes by Sex.
Let us begin our analysis by fitting a mixture of multivariate Gaussian distributions, using \textit{Mclust} package in R. The routine \textit{mclustBIC} fits a set of normal mixture models, on the base of the parameters you set in its call. We considered from 1 to 9 components in the mixture and different patterns for the covariance matrices, from the more constrained homoscedastic model, to the more general heteroscedastic one. After the estimation, \textit{mclustBIC} provides a model selection procedure based on the BIC, a well-known penalized likelihood criterium. In Figure \ref{fig:MisclAISdata}  the BIC values for each kind of model are shown, and for each choice of the number of mixture components. The  three letters in the acronym of the models stand respectively for the \textit{volume}, the \textit{shape} and \textit{orientation} of the ellipsoids of equal probability of the components, which could be Equal (hence E) or Variable (V) across the components. Notice that the shape may also be Isotropic (hence the letter I denotes spherical ellipsoids), in this case also the orientation is the same. Hence, we see that \textit{Mclust} suggests  an EEV model (ellipsoidal, equal volume and shape, different orientation of the component scatters)  with $G=2$ components, providing the highest BIC value, i.e. $BIC=-10251.6$.
%
Now, if we employ the model suggested by \textit{Mclust} to classify AIS data, we obtain 18 misclassified units, i.e., a misclassification error equal to $18/202=9.4\%$. In Figure \ref{fig:MisclAISdata} we can see the classification results.
Surely this is not an easy dataset for classification, due to the apparent class overlapping we saw in the first scatterplot in Figure \ref{fig:AISdata}.

To improve the classification, we may exploit the conjecture that a
strong correlation exists among the hematological and physical
measurement. Therefore we  fit  mixtures of factor analyzers,
assuming the existence of some underlying unknown factors (like
nutritional status, hematological composition, overweight status
indices, and so on) which jointly explains the observed
measurements.  Through the underlying factors, we aim at finding a
perspective on data which disentangle the overlapping components. To
avoid variables having a greater impact in the model (which is not
affine equivariant) 
 due to different scales, before
performing the estimation, the variables have been standardized to
have zero mean and unit standard deviation.
 We begin by adopting the  \textit{pGmm} package from R, that fits mixtures of factors analyzers with patterned covariances. Parsimonious Gaussian mixtures are obtained by constraining the loading $\Lambda_j$ and the errors $\Psi_j$ to be equal or not among the components.
We employed the routine \textit{pGmmEM}, considering
from 1 to 9 components, and number of underlying factors $d$ ranging
from 1 to 6, with 30 different random initialization, to provide the
best iteration (in terms of BIC) for each case. The best model is a
CUU mixture model with $d=4$ factors and $G=3$ components, with
$BIC=-3127.424$. CUU means Constrained loading matrices
$\Lambda_j=\Lambda$ and Unconstrained error matrices
$\Psi_g=\omega_g\Delta_g$, where $\Delta_g$ are normalized diagonal
matrices and $\omega_g$ is a real value varying across components.
Using this model to classify athletes,  we got 109 misclassified
units and we discarded it.

\begin{figure}
\begin{center}
    \includegraphics[height=8 cm, width=8 cm]{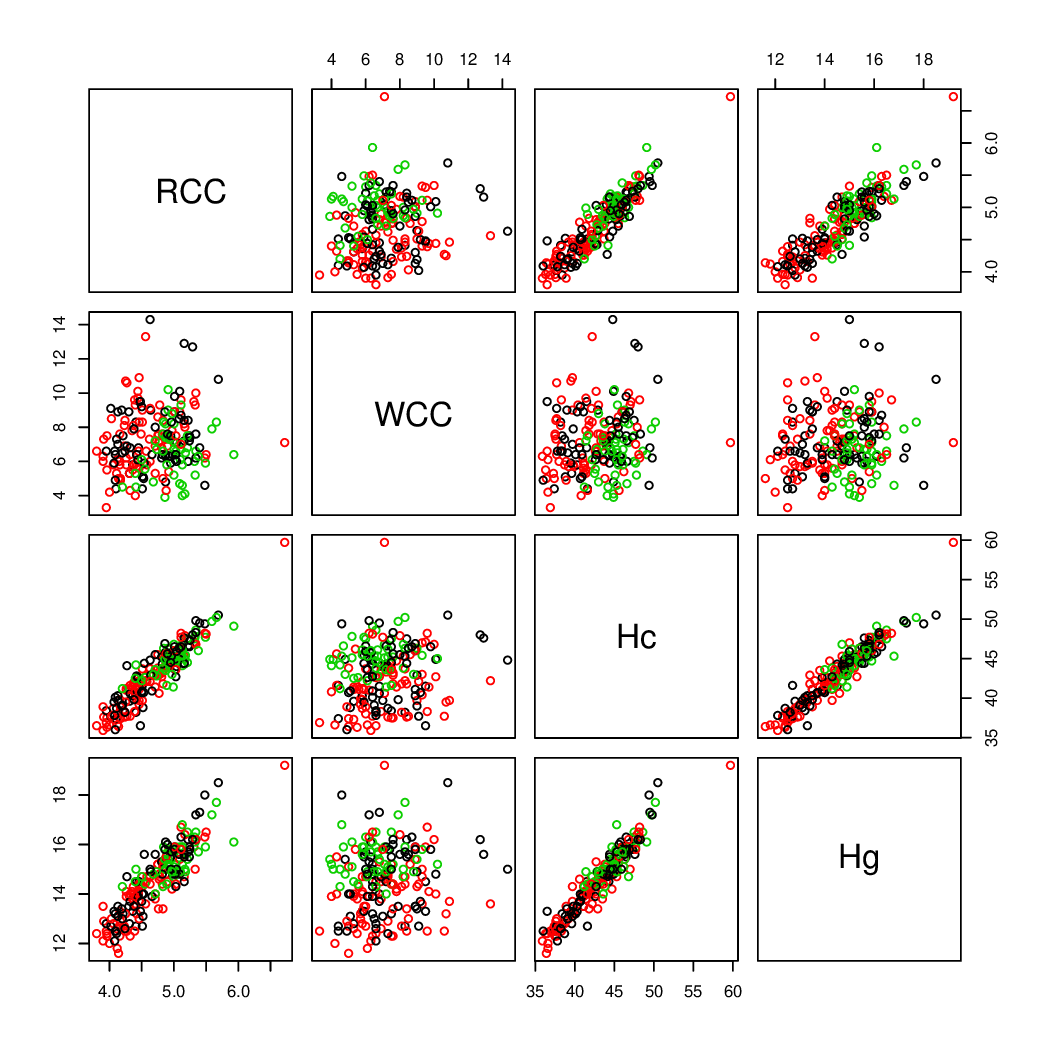}
\end{center}
\caption{\footnotesize The classification of AIS data obtained through the best UUU model from \textit{pGmm} with $G=2$ and $d=4$ (female data in red, male in green, misclassified units in black)}\label{fig:AIS_pGmm}
\end{figure}

As a second attempt using \textit{pGmm}, we estimated a UUU model by setting $G=2$ components, and $d=6$. The acronym UUU means that we leave the estimation of loadings $\Lambda_j$ and errors $\Psi_j$ unconstrained.
Based on 30 random starts, the best UUU model  had $BIC=-3330.306$, and the consequent classification of the AIS dataset produces 72 misclassifies units (misclassification error=35.6\%), that are visualized in Figure \ref{fig:AIS_pGmm}.

Finally, we want to show the performances of our  trimmed and constrained estimation for MFA on the AIS data. All the results are generated by the  procedure described in Section \ref{se3_2}, are based on 30 random initializations and returning the best obtained solution of the parameters, in terms of highest value of the final likelihood.

\medskip
\begin{table}[h!]
\centering
\footnotesize{\caption{\footnotesize Trimmed and constrained MFA estimation on the AIS data set (best results over 30 random initializations). Misclassification error $\eta$ (in percentage) under different settings \newline} 
\label{tab:AIS}       

\begin{tabular}{rcccccccc}
  \hline \hline
$c_{\rm noise}$                 &  $10^{10}$    & 45            &$10^{10}$      &  45       & $10^{10}$     & 45            & $10^{10}$     & 45  \\
$c_{\rm load}$              &  $10^{10}$        &  $10^{10}$    &  10       & 10            & $10^{10}$      & $10^{10}$    &  10       &  10  \\
 $\alpha$                   & 0             &  0            &  0            &  0            & 0.05      & 0.05      &  0.05               & 0.05 \\
 \hline
$\eta$          & 0.0891        & 0.1881        & 0.4554        & 0.0842        & 0.1039        & 0.1782        & 0.4505        &0.0149\\
\hline\hline
\end{tabular}
}
\end{table}
\medskip

\begin{figure}
\begin{center}
    \includegraphics[width=7.9 cm, height=8.5 cm]{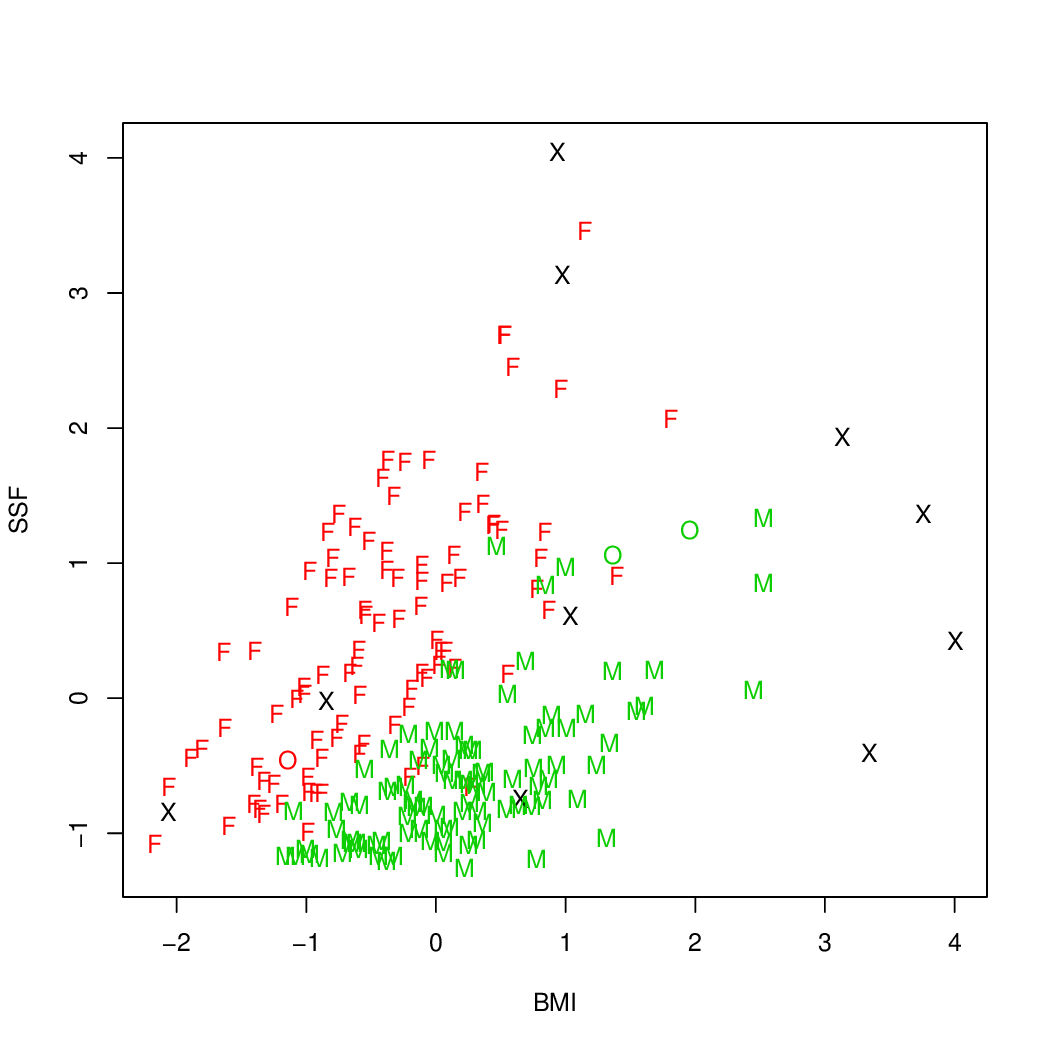}
    \includegraphics[width=7.9 cm, height=8.5 cm]{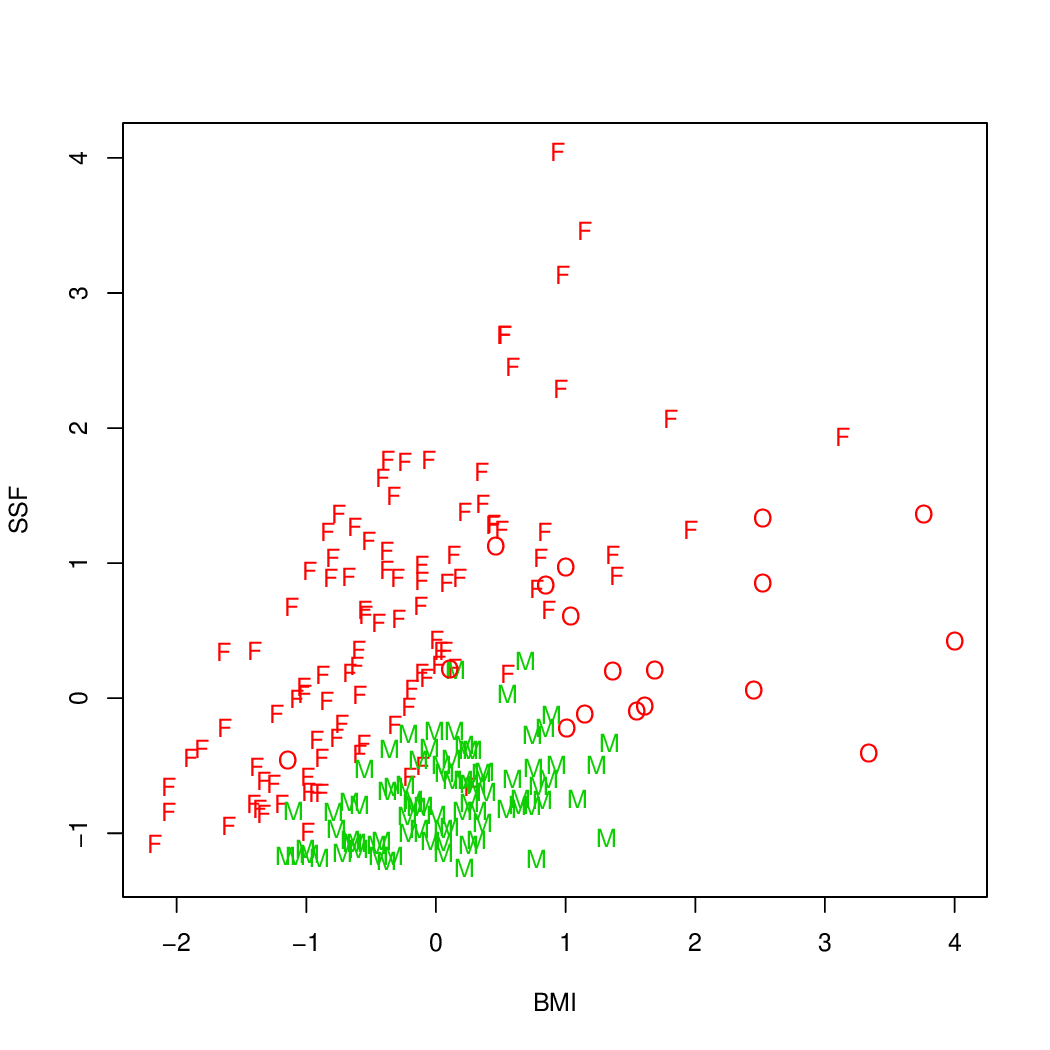}
\end{center}
\caption{\footnotesize Classification of AIS  data with fitted trimmed and constrained MFA (left panel), compared to non-robust MFA, i.e. the fitted model in 1st row of Table \ref{tab:AIS} (right panel). Misclassified data are denoted by O,  trimmed data by X.}
\label{fig:ResultsMFA}
\end{figure}

We see that the best solution, with only 3 misclassified points, has been obtained by combining trimming ($\alpha=0.05$) and constrained estimation of $\Psi_g$ ($c_{noise}=45$) and $\bLambda_g$ ($c_{load}=10$), with $d=6$. The choice of $d=6$ has been motivated by performing a factor analysis on the observations coming from the group of male athletes, in which we may employ the screw plot and  test the hypothesis that 6 factors are sufficient, with chi square statistic equal to 97.81 on 4 degrees of freedom, and  $p$-value$=2.88e-20$. In any case, results with $d \neq 6$ have also been checked. The constraints, and in first place the constraint on $\bPsi_g$, play
an important role (compare results in columns 2-4-6 and 8 to the ones displayed in the odd columns), but trimming is needed to reach the best result. This is motivated by the fact that the data, as a whole, are not following a 11-dimension multivariate Gaussian, as it can be easily checked by performing a previous Mardia test.
Two results of the fitted models and the subsequent classifications are displayed in Figure \ref{fig:ResultsMFA}, by selecting the 2 variables in the scatterplot that allow a better vision of trimmed and misclassified units. We have chosen to represent the best solution (upper panel), with only 3 misclassified points, denoted by ``O" in the graph, and with 10 trimmed points, denoted by ``X". In the lower panel, to make a comparison, we report classification results obtained by the fitted model in first row of Table \ref{tab:AIS}. In this second case, we were doing an almost unconstrained estimation of $\bPsi_g$ and $\bLambda_g$ and we were not applying trimming, obtaining 18 misclassified observations.


The misclassified observations are in rows 70,  73, and 121 in the AIS dataset. Two misclassified units are among female athletes, one is among male athletes.  The density of the mixture components  for the observation in position 73 are close (0.000021 and 0.00034), while for the other two observations they are neatly different.

\begin{figure}
\begin{center}
    \includegraphics[width=7.9 cm, height=8.5 cm]{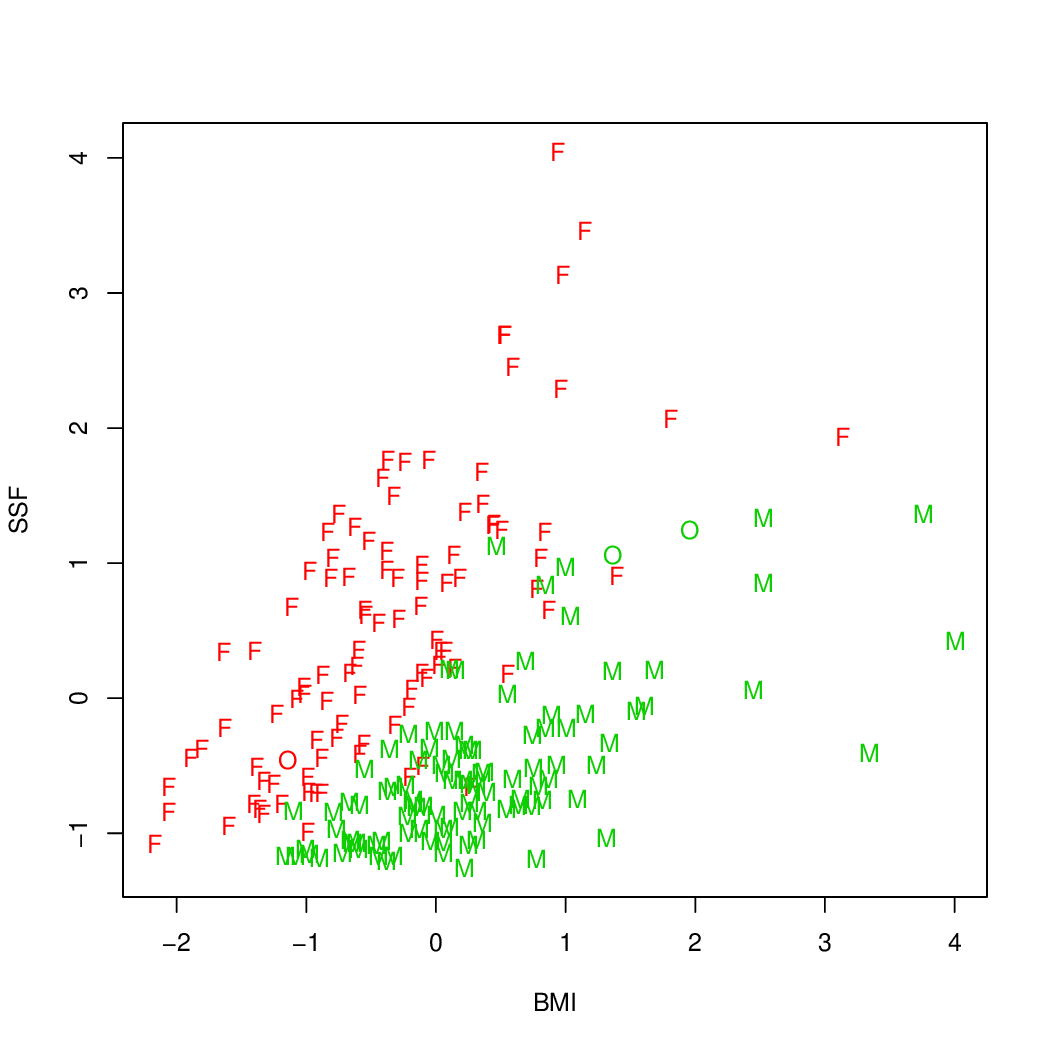}
\end{center}
\caption{\footnotesize Classification of AIS  data after classifying also trimmed observations. The three misclassified points are denoted by O, and they represent only $1.5\%$ of the data.}
\label{fig:AISfinal}
\end{figure}

Finally, we recall that trimmed observations were discarded to provide robustness to the parameter estimation. After estimating the model, hence, it makes sense to classify also these observations. The trimmed observations are in rows
 11,  56,  75,  93,  99, 133, 160, 163, 166, 178, and if we assign them by the Bayes rule to the component $g$ having greater value of  $D_{g}(\bx;\theta)=\phi_p \big(\bx ;\bmu_g,\bLambda_g \bLambda'_g+\bPsi_g \big)\pi_g$, we classify the first five in the female group of athletes, and the second group of five in the male group. This means that all the trimmed observations have been assigned to their true group.  Table \ref{tab:AISfinalclass} shows the details of the classification, and Figure \ref{fig:AISfinal} plots the final  result of the robust model fitting.


\medskip
\begin{table}[h!]
\centering
\footnotesize{\caption{\footnotesize Trimmed units in the AIS data set and their final classification \newline}
\label{tab:AISfinalclass}       

\begin{tabular}{cccc}
\\
  \hline \hline
unit     &$D_{1}(\bx;\theta)=\phi_p\big(\bx ;\mu_1,\bLambda_1\bLambda_1'+\bPsi_1\big)\pi_1$    &$D_{2}(\bx;\theta)=\phi_p\big(\bx ;\mu_2,\bLambda_2\bLambda_2'+\bPsi_2\big)\pi_2$ &Sex  \\
   \hline 
   
11      &1.4 e-15   &9.2 e-20   &F   \\
56  &7.2 e-08   &4.5 e-25   &F  \\
75  & 5.2 e-09  &1.2 e-11   &F   \\
93  &1.7 e-07   &1.0 e-10   &F  \\
 99     &1.2 e-09   &6.4 e-70   &F  \\
133     &9.8 e-85   &3.2 e-12   &M \\
160     &9.9 e-74   &1.5 e-08   &M \\
163     &9.9 e-87   &2.0 e-08   &M \\
166     &2.2 e-16   &1.4 e-13   &M \\
178 &3.1 e-23   &3.8 e-13   &M \\
\hline\hline
\end{tabular}
}
\end{table}
\medskip

As a last analysis on AIS dataset, we are interested in factor interpretation.
%
\medskip
\begin{table}[h!]
\centering
\footnotesize{\caption{\footnotesize Factor loadings in the AIS data set}
\label{tab:AISrot_factors}       
\begin{tabular}{lcccccc}
\\
  \hline
  \hline
  \\
&\multicolumn{6}{c}{rotated $\bLambda_1$ (female athletes)}\\

$RCC$ & $\boxed{0.697}$ & -0.006 &  -0.009&  -0.055 &  0.001&  -0.035\\
$WCC$ & 0.000 &  0.009&  0.000 & -0.015&   0.012 & $\boxed{-0.941}$\\
  $Hc$ &  $\boxed{0.794}$ & -0.015 &  0.040 & -0.004  & 0.010  & 0.026\\
$Hg$ &  $\boxed{0.682}$  & 0.021 & -0.025  & 0.047 & -0.002  & 0.007\\
  $Fe$ & 0.002& -0.005&  $\boxed{-0.510}$ &  0.003 & -0.004 &  0.000\\
  $BMI$ &  0.029 & -0.008 &  0.023  & $\boxed{0.644}$ & -0.316 & -0.057\\
$SSF$ & -0.040 & -0.012&  -0.037 &   0.033&  $\boxed{-0.889}$&  -0.017\\
$Bfat$ &  0.014 & -0.024&   0.013&  -0.007&  $\boxed{-0.826}$&   0.008\\
  $LBM$ &  0.022&  $\boxed{-0.419}$&   0.020 &  0.295&   0.054 & -0.025\\
  $Ht$ &  0.020 & $\boxed{-0.924}$ &  0.023 & -0.128&  -0.076&  -0.002\\
 $ Wt$  & 0.029 & $\boxed{-0.468}$&   0.023&   0.330 & -0.235&  -0.031\\
\\
&\multicolumn{6}{c}{ rotated $\bLambda_2$  (male athletes)}\\
$RCC$ &-0.033 & $\boxed{0.663}$ & 0.077 &-0.015 &-0.025 &-0.053\\
$WCC$ &  0.003 &-0.004 & 0.024 &-0.003&  $\boxed{1.024} $& 0.013\\
  $Hc$ & 0.048 & $\boxed{0.622}$& -0.008 & 0.005 & 0.036 &-0.048\\
$Hg$ & -0.002& $\boxed{0.604}$ &-0.051 & 0.009 & 0.001 & 0.079\\
  $Fe$ & 0.010 & -0.006 & 0.027 &$\boxed{1.103}$ & -0.004 & 0.008\\
 $BMI$ & -0.371 & 0.109 &$\boxed{-0.656}$ & 0.074 & 0.070 &-0.261\\
$SSF$ &$\boxed{-0.616}$& 0.002 & 0.009 &-0.001& 0.015 &-0.026\\
$Bfat$ & $\boxed{-0.610}$ &-0.009 & 0.003 & 0.007 &-0.000 & 0.040\\
 $ LBM$ &  0.036 & 0.071 &-0.344 & 0.037 & 0.053 &$\boxed{-0.885}$\\
  $Ht$ &  0.036 & 0.005 & 0.170 & -0.022 & 0.009 &$\boxed{-1.157}$\\
  $Wt$ & -0.222 & 0.071 &-0.357 & 0.042 & 0.056& $\boxed{-0.884}$\\
\\ \hline \hline
\end{tabular}
}
\end{table}
\medskip
\normalsize

The rotated factor loading matrices have been obtained by employing a Gradient Projection algorithm, available through the R package \textit{GPArotation} \citep{bernaards2005gradient, browne2001overview}. We opted for an oblimin transformation, which yielded results shown in Table \ref{tab:AISrot_factors}. We observe that the two groups highlight the same factors, while in a slightly different order of importance. The first factor for the group of observations for female athletes, may be labelled as a \textit{hematological factor}, with a very high loading on $Hc$, followed by $RCC$ and $Hg$. The second factor, loading heavily on $Ht$, and in a lesser extent on $Wt$ and $LBM$, may be denoted as a \textit{general nutritional status}. The third and the fourth factors are related only to $Fe$ and $BMI$, respectively. The fifth factor can be viewed as a \textit{overweight assessment index} since $SSF$ and $Bfat$ load highly on it. The sixth factor is related only to $WCC$. Noticing that $WCC$ is not joined into the \textit{hematological factor}, we observe that the specific role of lymphocytes,  cells of the immune system that are involved in defending the body against both infectious disease and foreign invaders, seems to be pointed out.
Analogous comments may be done on the factor loadings for the group of male athletes.

\section{Concluding remarks}\label{sec:Concluding}
In this paper we propose a robust estimation for the mixture of
Gaussian factors model. To resist pointwise contamination and sparse
outliers that could arise in data collection, we adopt and
incorporate a trimming procedure along the iterations of the EM
algorithm. The key idea is that a small portion of observations
which are highly unlikely to occur, under the current fitted model
assumption, are discarded from contributing to the parameter
estimates. Furthermore, to reduce spurious solutions and avoid
singularities of the likelihood,  we implement a constrained ML estimation for
the component covariances. Results from Monte Carlo experiments show
that bias and MSE of the estimators in several cases of contaminated
data are comparable to results obtained on data without noise.
Finally, the analysis on a real dataset illustrates that robust
estimation leads to better classification and provides direct
interpretation of the factor loadings.

Further investigations are needed to
tune the choice of the parameters, such as the portion of trimming data and the values of the constraints.
Though interesting, this issue is beyond the scope of the present paper. Surely, the
researcher may specify the partial information he may have about the
shape of the expected clusters from the data at hand, hence providing a part of
these parameters. Then,  data-dependent diagnostic based on
trimmed BIC notions \citep{NeyF07} and/or graphical tools such as the ones  in \cite{GarG11}, conveniently adapted to the specific case, could assist in taking appropriate choices for the rest of
the parameters.  The encouraging results here obtained suggest a
deeper discussion of these implementation details in a future work.
\section*{Acknowledgements}
\footnotesize
This research is partially supported by the Spanish Ministerio de Ciencia e Innovaci\'{o}n, grant MTM2011-28657-C02-01, by Consejer\'{i}a de Educaci\'{o}n de la Junta de Castilla y Le\'{o}n, grant VA212U13, and by grant FAR 2013 from the University of Milano-Bicocca.






\end{document}